\documentclass[showpacs,amsmath,amssymb,aps,10pt,twocolumn]{revtex4-2}
\usepackage[utf8]{inputenc}
\usepackage{bbold}
\usepackage[margin=0.9in]{geometry}
\usepackage{graphicx}
\usepackage{braket}
\usepackage{fixmath}
\usepackage{physics}
\usepackage[english]{babel}
\usepackage{amssymb}
\usepackage{amsmath}
\usepackage{dsfont}
\usepackage{multirow}
\usepackage{tabularx}
\usepackage{units}

\usepackage{xspace}
\usepackage{dsfont}
\usepackage{bm}

\usepackage[pdfencoding=auto, psdextra]{hyperref}

\newcommand{\mycomment}[1]{}

\usepackage[usenames, dvipsnames]{xcolor}

\begin{document}

\title{Correlated quantum dynamics of graphene}

\author{Fran{\c c}ois Rousse}
\affiliation{School of Science and Technology, \"{O}rebro University, 70182 \"{O}rebro, Sweden }

\author{Olle Eriksson}
\affiliation{School of Science and Technology, \"{O}rebro University, 70182 \"{O}rebro, Sweden }
\affiliation{Division of Materials Theory, Department of Physics and Astronomy, Uppsala University, Uppsala, Sweden }
\author{Magnus \"Ogren}
\affiliation{School of Science and Technology, \"{O}rebro University, 70182 \"{O}rebro, Sweden }
\affiliation{Hellenic Mediterranean University, P.O. Box 1939, GR-71004, Heraklion, Greece}



\date{\today}


\begin{abstract}
Phase-space representations are a family of methods for dynamics of both bosonic and fermionic systems, that work by mapping the system's density matrix to a quasi-probability density and the Liouville-von Neumann equation of the Hamiltonian to a corresponding density differential equation for the probability. 
We investigate here the accuracy and the computational efficiency of one approximate phase-space representation, called the fermionic Truncated Wigner Approximation (fTWA), applied to the Fermi-Hubbard model. On a many-body 2D system, with hopping strength and Coulomb $U$ tuned to represent the electronic structure of graphene, the method is found to be able to capture the time evolution of first-order (site occupation) and second-order (correlation functions) moments significantly better than the mean-field, Hartree-Fock method. The fTWA was also compared to results from the exact diagonalization method for smaller systems, and in general the agreement was found to be good. The fully parallel computational requirement of fTWA scales in the same order as the Hartree-Fock method, and the largest system considered here contained 198 lattice sites.

\end{abstract}
\maketitle
\section{Introduction}

Since its discovery \cite{novoselov2004electric}, graphene has become an extremely rich arena for research, both from fundamental science as well as for practical applications \cite{katsnelson2007graphene, geim2010rise}. The electronic structure of graphene (and graphite) predates the experimental observation of graphene with a broad margin \cite{wallace1947band, mcclure1956diamagnetism}, where the initial efforts were based on tight-binding electronic structure theory. Since these early efforts many studies have been published (see e.g. Ref. \cite{neto2009electronic}), with results of the electronic structure that do not deviate significantly from the early results \cite{wallace1947band, mcclure1956diamagnetism}. One of the most interesting aspects of the electronic structure of graphene is the linear dispersion relation around the so called Dirac point, $K$, at the Brillouin zone boundary, with its peculiar consequences for Klein tunneling \cite{katsnelson2006chiral,katsnelsonbook} and half-integer quantum Hall effect \cite{novoselov2004electric}. The unique electronic properties of graphene around the Fermi level has opened for possible applications in electronics and spintronics (see e.g. Refs. \cite{bolotin2008ultrahigh, panda2020ultimate}). 

The valence band states that have attracted most attention are the so called $\pi$ and $\pi^*$ states, that represent occupied and unoccupied electron states of undoped graphene. These states are composed of $p_z$ orbitals centered at each C atom. These states are weakly bonding, and the strong chemical bonds of graphene come instead from $sp^2$ hybrids (composed of $s$, $p_x$ and $P_y$ orbitals of each C atom) that build up a strong network of $\sigma$ bonds (see e.g. Ref. \cite{harrison2012electronic}). The energy bands corresponding to these $\sigma$ bonds are however far below the Fermi level, and are from a transport point of view rather uninteresting. This produces a rather interesting scenario, where the basic electronic structure close to the Fermi level (a few electron volts on either side of the Fermi level) of the bipartite graphene system can be described by tight-binding theory with one orbital ($p_z$) per atomic site. 
We will utilize the simplicity of the orbital structure of the $\pi$ and $\pi^*$ states in this paper, by investigating the electronic structure using tight-binding theory, including on-site correlations, as described by the Hubbard model (see Section \ref{microscopic}). This allows to study a typical many-body, model Hamiltonian, with the advantage of its ability to describe a real, physical system. 

The dynamical properties of the electronic structure of graphene is the main focus of this investigation, and we have employed several approximations to do this, they are: the Hartree-Fock (HF) method, Exact Diagonalisation (ED) as well as the fermionic Truncated Wigner Approximation (fTWA)\cite{sajna2020semiclassical}. Of these the ED method is exact but only tractable for small systems because of its exponential scaling, while the other two represent approximations described and analyzed below. It is noteworthy that the fTWA method is the least frequently investigated approximation, when it comes to electronic structure theory, and we will for this reason put special emphasis on this method. The basic equation describing the quantum dynamics here is the Liouville-von Neumann equation of the density matrix, and we detail below the different technical aspects of its solution, comparing in particular the time evolution of the occupation numbers and second-order correlation functions.

\section{The microscopic Hamiltonian}
\label{microscopic}

In this work, we consider the Hubbard-Fermi Hamiltonian, written in the second quantization formalism:
\begin{equation}
    \label{eq:FHeq1}
    \hat{H} = -  \sum_{  i,j , \sigma} j_{ij} \hat{c}_{i \sigma}^{\dagger} \hat{c}_{j \sigma} +  \sum_{i,j} u_{ij}  \hat{c}_{i \uparrow}^{\dagger} \hat{c}_{i \uparrow} \hat{c}_{j \downarrow}^{\dagger}   \hat{c}_{j \downarrow} ,
\end{equation}
where $j_{ij}$ is the hopping interaction of $p_z$ orbitals between sites $i$ and $j$, $\sigma$ is the particle's spin, while $u_{ij}$ is the interaction between two particles of opposite spins situated on sites $i$ and $j$. Furthermore, $\hat{c}$ is an annihilation operator and $\hat{c}^{\dagger}$ a creation operator. In this article we only numerically  consider hopping between neighbour sites and with equal potentials: $j_{ij} = J$ if $i$ and $j$ are neighbours, and in addition we limit the work to a system with on-site interaction: $u_{ij} = U$  if $i=j$. In the present work we chose equal strengths of $J$ and $U$.
The Hamiltonian~(\ref{eq:FHeq1}) is a good representation of the electronic structure of graphene around the Fermi level, according to the discussion of the introduction. Hence this Hamiltonian has been used in several instances to simulate the energy dispersion of a graphene layer (see e.g. Refs. \cite{ma2011magnetic} and \cite{joost2019femtosecond}).

\section{Fermionic Truncated Wigner Approximation}

Phase-space representations are a family of methods that have already demonstrated their ability to model the dynamic of many-body bosonic systems. They work by mapping the system's density matrix to a quasi-probability density and the Liouville-von Neumann equation of the Hamiltonian to a corresponding density differential equation for the probability. More recently, phase-space methods have been adapted to model fermionic dynamics. They are especially useful for 2D and 3D systems for which DMRG (density matrix renormalisation group), and alike methods, are less successful. 
We focus here on the computational efficiency of one approximate phase-space representation, called the fermionic Truncated Wigner Approximation (fTWA), and apply it to the Fermi-Hubbard model Eq.~(\ref{eq:FHeq1}).

We start by providing a brief introduction to fTWA. In phase-space representations, we formulate the problem via an expansion of the density operator $\hat{\rho}$ over an over-complete operator basis $\hat{\Lambda}(\mathbold{\lambda})~$\cite{walls2007quantum,corney2006gaussian, ogren2010first}:

\begin{equation}
    \hat{\rho}(t) = \int W(\mathbold{\lambda},t) \hat{\Lambda}(\mathbold{\lambda}) d \mathbold{\lambda} \text{  ,}
    \label{eq:w}
\end{equation}
where the expansion ‘coefficients’ $W (\mathbold{\lambda})$ constitute a quasi-distribution over generalised complex phase-space variables $\mathbold{\lambda}$. The Liouville-von Neumann equation that describes the density operator dynamic is then mapped into a partial differential equation (PDE) of the Wigner function $W$ (for more details see Appendix~\ref{sec:Motion_eq_ftwa}). In the Wigner-Weyl representation, phase-space variables are mapped to symmetrized operators, which leads to a PDE containing only odd-order derivatives. In particular, the PDE for the Wigner function has no diffusion term (second-order derivative). The time dependent quantum operators $\hat{\mathcal{O}}(t)$ are mapped to their Weyl symbols $\mathcal{O}_W$ and evaluated in the Heisenberg representation \cite{davidson2017semiclassical}:
\begin{equation}
    \langle \hat{\mathcal{O}}(t) \rangle = \int W(\mathbold{\lambda}, t) \mathcal{O}_W(\mathbold{\lambda} )  d \mathbold{\lambda}\text{  .}
    \label{eq:obs}
\end{equation}
 
The Truncated Wigner Approximation (TWA) is the practical implementation of the Wigner method. It here relies on two approximations: The high-order derivative terms (third-order and above) of the PDE are truncated, and the initial density is chosen to represent the two first moments (average and covariance). We use a gaussian distribution, similar to what was done in Ref.~\cite{davidson2017semiclassical}. Hence the Wigner function dynamics is found by computing trajectories of deterministic differential equations whose initial conditions are drawn from a gaussian probability density.

Because of the anti-commuting property of fermionic operators, they cannot directly be represented by complex numbers, so to adapt the Wigner representation to fermions, we choose to map phase-space variables to bilinear operators~\cite{davidson2017semiclassical, sajna2020semiclassical}: 

\begin{equation}
\begin{split}
    \hat{E}_{i \sigma_i}^{j \sigma_j} &\equiv \frac{1}{2}(\hat{c}_{j \sigma_j}^{\dagger} \hat{c}_{i \sigma_i} - \hat{c}_{i \sigma_i} \hat{c}_{j \sigma_j}^{\dagger} )     \text{ ,     } \\
        \hat{E}^{j \sigma_j, i \sigma_i} \equiv &\hat{c}_{j \sigma_j}^{\dagger} \hat{c}_{i \sigma_i}^{\dagger}\text{ , } ~~ \hat{E}_{j \sigma_j, i \sigma_i} \equiv \hat{c}_{j \sigma_j} \hat{c}_{i \sigma_i} \text{ .} 
    \end{split}
    \label{eq:ps_var}
\end{equation}
Here, letters $i$ and $j$ label site indices and $\sigma_i$ labels the particle spin. In a system with constant particle number, we only use $\hat{E}_{j \sigma_j}^{i \sigma_i}$, and we call $\rho_{i  \sigma_i, j \sigma_j}$ its corresponding complex phase-space variable. We also consider in this article that electrons will not flip their spin, which is a reasonable assumption since the spin-orbit coupling in graphene is very weak~\cite{katsnelsonbook}. For this reason we can limit the representation to only same-spin phase-space variables $\rho_{i j \sigma}$. The observable values are recovered using their Weyl symbol and Eq.~(\ref{eq:obs}), for example, the occupation operator and the doublon operators are linked to statistical averages of phase-space variables:

\begin{equation}
\begin{split}
\langle \hat{c}_{i\sigma}^{\dagger} \hat{c}_{j\sigma}\rangle &= \overline{ \rho_{ij\sigma}}  + \delta_{ij} / 2  \text{  ,} \\
 \langle \hat{c}_{i\sigma}^{\dagger} \hat{c}_{j\sigma}^{\dagger} \hat{c}_{j\sigma} \hat{c}_{i\sigma}\rangle &= \overline{\rho_{ii\sigma} \rho_{jj\sigma}} + \frac{1}{2}(\overline{ \rho_{ii\sigma}} + \overline{ \rho_{jj\sigma} }  )+ \frac{1}{4}\text{ ,}
\label{eq:fTWA_moments}
\end{split}
\end{equation}
see Appendix~\ref{sec:Motion_eq_ftwa} for details. The line above the symbols in Eq.~(\ref{eq:fTWA_moments}) denotes stochastic averages. The time evolution of the quantum system is then given by a first order PDE for $W$, which provides differential equations for $\rho_{i j\sigma}$ \cite{davidson2017semiclassical}:

\mycomment{
\begin{equation}
   i \frac{d}{dt} \rho_{i j\sigma} = \sum_{k} \rho_{ik \sigma} (u_{kj} \rho_{kk\bar{\sigma}}  - j_{kj}   )- (u_{ik}  \rho_{ii\bar{\sigma}}  - j_{ik}  ) \rho_{kj \sigma}  \text{ ,}\\
\end{equation}
}

\begin{equation}
\begin{split}
       \frac{\partial }{\partial t} \rho_{ij \sigma} &=  i \sum_{ k}   \bigg( j_{j k}  \rho_{i k \sigma}  -   j_{k i} \rho_{kj\sigma}  \\
       &+(u_{ki} -u_{k j}) \Big(\rho_{kk \bar{\sigma}} +\frac{1}{2} \Big) \rho_{i j \sigma} \bigg) \text{,}
       \label{eq:drho}
 \end{split}
\end{equation} 
where the hopping, $j_{j k}$, and Coulomb repulsion, $u_{ki}$, are defined in Eq.~(\ref{eq:FHeq1}).
From a practical point of view, $W(\rho, t)$ is represented by a set of independent realizations of Eq.~(\ref{eq:drho}), called trajectories, whose initial condition are distributed respecting the moments in~(\ref{eq:fTWA_moments}).

If the initial condition is a thermal state in a diagonal basis with site occupations $n_{i i \sigma}$, we can compute the first and second moments (mean and covariance) between phase-space variables to generate an initial gaussian density with the same moments, which gives:
\begin{equation}
\begin{split}
\text{if } i=j ~~~& \rho_{ii\sigma}(0) = n_{ii\sigma}(0) - \frac{1}{2}\text{ , }   \\
\text{and if} \\
i\neq j~~~& \rho_{ij\sigma}(0) =  \xi_{ij \sigma}  \sqrt{\frac{n_{ii \sigma} + n_{jj \sigma} - 2 n_{ii \sigma} n_{jj \sigma}}{2}} \text{  ,}
\end{split}
\label{eq:ic_fTWA}
\end{equation}
where $\xi_{ij \sigma}$ is from a complex normal distribution, with $\xi_{ji \sigma} = \xi_{ij \sigma}^*$. Again, we outline details of the calculations further in Appendix~\ref{sec:Motion_eq_ftwa}.

In this article the Hamiltonian~(\ref{eq:FHeq1}) is time-independent. However, time-dependence in e.g. an external potential is straightforward to implement in fTWA, and appears as explicit time-dependent terms in Eq.~(\ref{eq:drho}), just as they appear in the corresponding mean-field method (below). We have numerically evaluated fTWA against the other computational methods used in this article also for time-dependent Hamiltonians (not presented here).

\subsection{Exact Diagonalisation and Hartree-Fock}

For the evaluation of the ED dynamics, we compute step by step the general solution of the time-dependent Schr{\"o}dinger equation,
\begin{equation}
\ket{\Psi(t)} = \ket{\Psi_0}e^{\frac{i}{\hbar}\hat{H}t} \text{ , }
\label{eq:shro}
\end{equation}
using the Krylov subspace projection technique implemented in the expokit package~\cite{sidje1998expokit}. For details of these calculations we refer to Appendix~\ref{sec:edb}.

To obtain the mean-field dynamics, we use the Heisenberg equation of motions to compute the dynamic of the number operators $\hat{n}_{ij\sigma}$. Then we map their factorised averages to the variables of the mean-field method $\langle \hat{n}_{ij\sigma} \rangle \rightarrow n_{ij\sigma}$, see Appendix~\ref{sec:Motion_eq} for details. We recover the same differential equations that are found for the phase-space variables $\rho_{ij\sigma}$, as shown in~\cite{rahav2009gaussian}. Without the initial noise ($ \xi_{ij \sigma} = 0 $), Eqs.~(\ref{eq:drho}) and (\ref{eq:ic_fTWA}) provide the same results as the Hartree-Fock mean-field method.

\section{Results}


Below we present the results for two examples, we start with a small systems where a comparison between different theoretical methods can be made (Exact Diagonalization, fTWA  and Hartree-Fock), then we study a larger system on which the Exact Diagonalisation method is unable to give a result.

\subsection{Small graphene systems}

\begin{figure}
\begin{center}
\includegraphics[height=6 cm]{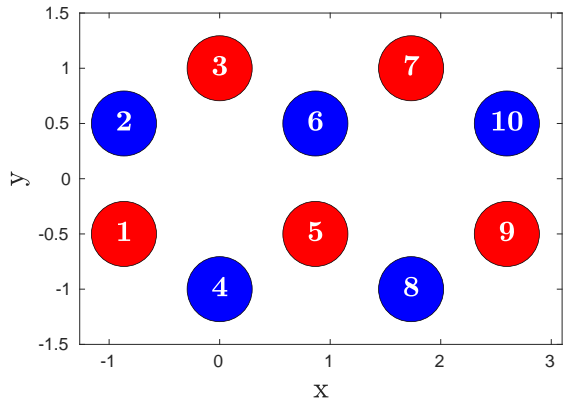}
\caption{(Color online) Illustration of one of the two dominating pure states of the ground state of the Hamiltonian~(Eqn.\ref{eq:FHeq1}). Here we consider a 10 sites Fermi-Hubbard model with half filling. The geometry is composed of two perfect hexagons of equal side lengths. This state was used as initial condition for the quantum dynamics calculations. In blue, the spin-down sites, in red the spin-up sites. The second most dominating state is the spin symmetry of this one, i.e. a state where all electrons have flipped their spin (and blue and red colours have been interchanged in the figure).}
\label{fig:10p_state1}
\end{center}
\end{figure}

We first study the accuracy of the fTWA method on a small graphene-like few-body system. The system is composed of ten sites organised in two hexagonal cells, see Fig.~\ref{fig:10p_state1}, which makes it large enough to be interesting and small enough to compute numerical solutions with the Exact Diagonalisation method. The system is assumed to be electronically half-filled, which means that there are as many particles as there are sites. We also consider a system with equal amount of spin-up as spin-down electrons. For the initial condition, we choose the $p$-particles Fock-space vector with the largest overlap to the ground state, where each site is filled with either a spin-up or spin-down particle, see Fig.~\ref{fig:10p_state1}. We motivate this choice further in Appendix~\ref{sec:gs}.
The time evolution with the fTWA method has been computed with $10^5$ trajectories, and until $t=5$, a choice that was made so that one can see when all methods considered here start to deviate from each other.  

\subsection{Evolution of occupations, first-order moments }

\begin{figure}
\begin{center}
\includegraphics[height=6.5 cm]{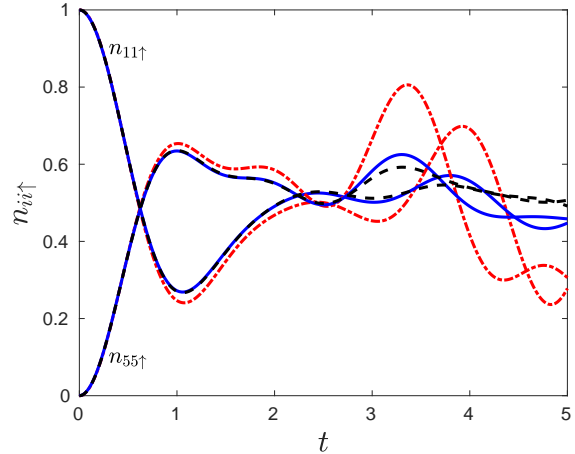}
\caption{(Color online) Dynamics of site occupations for spin-up particles on sites 1 ($n_{11}$) and 5 ($n_{55}$). The solid (blue) curves show the fTWA dynamics, the dashed (black) ones show the exact diagonalisation dynamics, the dashed-dotted (red) ones the Hartree-Fock dynamics. For the occupation, the HF dynamic starts to deviate from the ED at $t\simeq1$, while the fTWA dynamic starts to deviate from the ED solution at around $t\simeq3$.}
\label{fig:10p_occup}
\end{center}
\end{figure}

The first measure used to compare the methods is the time evolution of site occupations, $n_{ii\sigma} (t)$. We show in Fig.~\ref{fig:10p_occup} the occupation of spin-up particles on sites $1$ and $5$, respectively $n_{11\uparrow}$ and $n_{55\uparrow}$. We see that the occupations computed with the Hartree-Fock method deviate from the exact solution (the ED method) at $t\simeq 1$ and that the occupations computed with fTWA starts to deviate later, at $t \simeq 3$. We stress that the only computational cost to achieve this improvement is that the calculation is repeated, in parallel, for the different trajectories.

Note that the computation times of both HF and fTWA methods scale quadratically with the system size, and the fTWA method is here correct for approximately three times longer. 
Here, using $10^5$ trajectories in the fTWA method, the statistical error is negligible, hidden behind the width of the line. Hence, the deviation comes entirely from the truncation possible from the approximation of the initial density~\cite{davidson2017semiclassical} in the formalism of the method, and is not due to statistical uncertainty. 
We also note an improvement on the long term dynamics. When $t\rightarrow \infty$ the occupation stabilizes at $n_{ii \sigma} \simeq 0.5$ for all sites and spins (data not shown). This result is recovered by the fTWA method, but not by the HF one that oscillates uncontrollably.   

\subsection{Evolution of correlations, second-order moments}

One advantage of phase-space methods, like fTWA, is their ability to give   information on the dynamics of higher-order moments, like second-order correlation functions, even for large system. The correlation function between two sites $i$, $j$ and spins $\sigma_1$, $\sigma_2$, here denoted $g^{(2)}_{i \sigma_1, j \sigma_2}$, can be seen as the effect the presence of a spin-$\sigma_1$ particle on site $i$ has on the probability to have a spin-$\sigma_2$ particle on site $j$. 
 The explicit formulas are, in the Schr\"{o}dinger picture for ED:
\begin{equation}
    \label{eq:g2_ED}
    g^{(2)}_{i \sigma_1, j \sigma_2 } (t)= \frac{\langle  \hat{c}_{i \sigma_1}^{\dagger}\hat{c}_{j \sigma_2}^{\dagger} \hat{c}_{i \sigma_1}\hat{c}_{j \sigma_2} \rangle }{\langle \hat{c}_{i \sigma_1}^{\dagger} \hat{c}_{i \sigma_1}\rangle \langle \hat{c}_{j \sigma_2}^{\dagger} \hat{c}_{j \sigma_2}\rangle} \text{  ,}
\end{equation}
and in the Heisenberg picture with the fTWA phase-space variables:

\mycomment{
\begin{equation}
    \label{eq:g2_fTWA}
    g^{(2)}_{i \sigma_1, j \sigma_2 } (t)= \frac{\langle \hat{n}_{ii \sigma_1} \hat{n}_{jj \sigma_2} \rangle}{\langle \hat{n}_{ii \sigma_1} \rangle \langle \hat{n}_{jj \sigma_2} \rangle} \text{  .}
\end{equation}
}

\begin{equation}
\begin{split}
    \label{eq:g2_fTWA}
    g^{(2)}_{i \sigma_1, j \sigma_2 } (t)&= \frac{\overline{\rho_{ii\sigma} \rho_{jj\sigma}} + (\overline{ \rho_{ii\sigma}} + \overline{ \rho_{jj\sigma} }  ) / 2+ 1/4} {(\overline{ \rho_{ii\sigma}}  + 1 / 2) (\overline{ \rho_{jj\sigma}}  + 1/ 2 )} \\
    &= \frac{\overline{( \rho_{ii\sigma} + 1 / 2) (\rho_{jj\sigma}  + 1/ 2 )}} {(\overline{ \rho_{ii\sigma}}  + 1 / 2) (\overline{ \rho_{jj\sigma}}  + 1/ 2 )}  \text{  .} 
    \end{split}
\end{equation}
For the mean-field Hartree-Fock approximation we use Wick's theorem, see Eq.~(\ref{eq:wickthm}) in Appendix~\ref{sec:Motion_eq_hf}. 

In Fig.~\ref{fig:10p_g2_124}, we plot the correlation between the spin-up particle on site $1$ and the spin-up particles on the two neighboring sites, $2$ and $4$, see Fig.~\ref{fig:10p_state1} for the geometry. The correlation computed with fTWA is essentially the same as that of ED, until approximately $t \simeq 2$ where it starts to deviate visibly. This happens earlier than for the occupation, shown in Fig.~\ref{fig:10p_occup}. At $t \simeq 0$, we observe that the correlation functions computed with fTWA does not tend precisely to its theoretical value, Eq.~(\ref{eq:form_g_st}), of $g^{(2)}_{1 \uparrow, 2 \uparrow} = 0.5$. The quantitative difficulties fTWA has to compute correlations involving initial empty sites are known for bosons~\cite{sinatra2002truncated}, the calculations made in Appendix~\ref{sec:asymptotes} expose the same problem for the present Hamiltonian. In principle, these results can be improved by adding more trajectories or using projection methods.

\begin{figure}
\begin{center}
\includegraphics[height=6.5 cm]{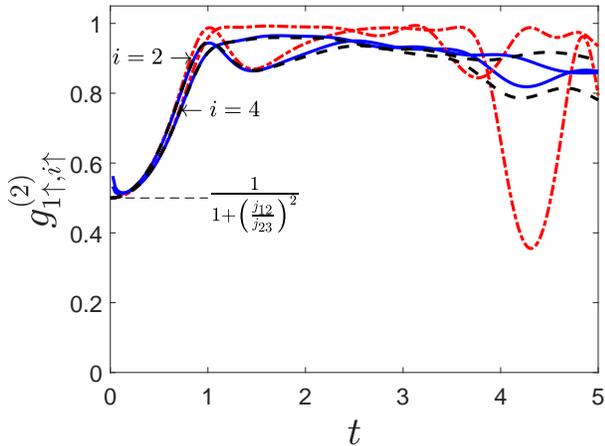}
\caption{(Color online) Dynamics of correlation functions between neighbour sites, here $g^{(2)}_{1 \uparrow, 2 \uparrow }$ and $g^{(2)}_{1 \uparrow, 4 \uparrow }$. The different curves represent the same methods and parameter values as described in Fig.~\ref{fig:10p_occup}. 
}
\label{fig:10p_g2_124}
\end{center}
\end{figure}

\begin{figure}
\begin{center}
\includegraphics[height=6.5 cm]{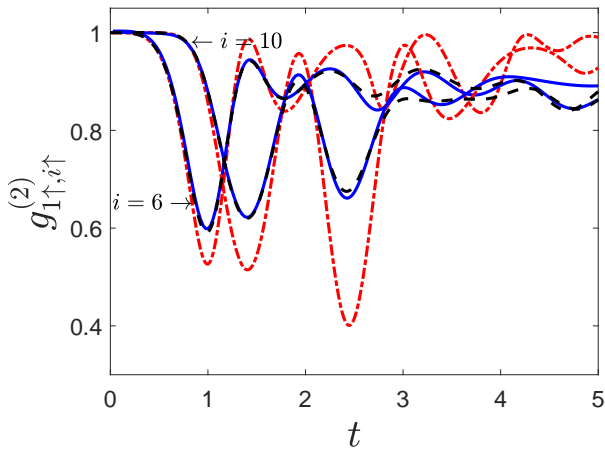}
\caption{(Color online) Dynamics of correlation functions between distant sites, here $g^{(2)}_{1 \uparrow, 6 \uparrow }$ and $g^{(2)}_{1 \uparrow, 10 \uparrow }$. The different curves represent the same methods and parameters values as described in Fig.~\ref{fig:10p_occup}. 
}
\label{fig:10p_g2_1610}
\end{center}
\end{figure}

In Fig.~\ref{fig:10p_g2_1610}, we plot the correlations between the spin-up particle on site $1$ and the spin-up particles on distant sites, $6$ and $10$, see again Fig.~\ref{fig:10p_state1} for the geometry. We observe that the fTWA results starts to deviate from data obtained by ED, at a time between $t \simeq 2$ and $t \simeq 3$. We also observe in Fig.~\ref{fig:10p_g2_1610} that the time for a correlation to appear between two sites, i.e. when $g^{(2)}$ deviate from unity, depends on the increasing  distance between those two sites. 

In the case of different spins correlations, like e.g. $g^{(2)}_{i\uparrow, i \downarrow}$, the second term in Eq.~(\ref{eq:wickthm}) is zero. As a consequence HF gives a constant, $g^{(2)} \equiv 1$, and cannot be used for comparisons. However, fTWA gives accurate results for short times, as we have explored numerically in comparisons with ED for small systems.

\subsection{Large-time correlation functions}

\begin{figure}
\begin{center}
\includegraphics[height=6.5 cm]{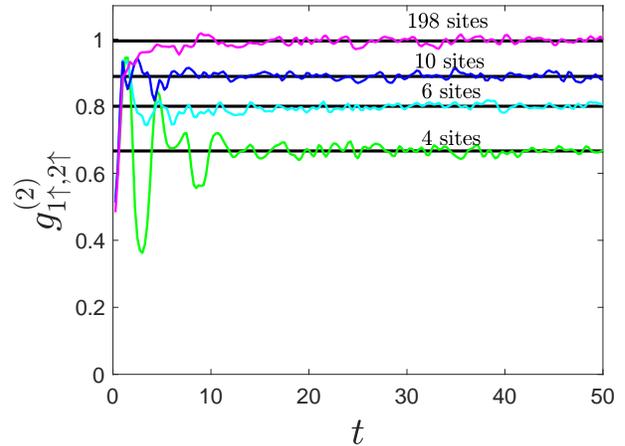}
\caption{(Color online) Long-time fTWA dynamics for correlations between spin-up particles on sites $1$ and $2$, $g^{(2)}_{1 \uparrow, 2 \uparrow} $, for systems of different size, $n=4,6,10, 198$. The horizontal (black) lines represent the limit ($t \rightarrow \infty$) values of correlation function gathered in table~\ref{tab:g2_res}, we recover the values of formula~(\ref{eq:form_g_lt_2}) for all cases.} 
\label{fig:np_g2}
\end{center}
\end{figure}

To investigate if the fTWA is also able to model the long term values of correlation functions, we computed the fTWA results of $g^{(2)}_{1 \uparrow, 2 \uparrow}$ for systems of different size, $n=4,6,10, 198$ sites, until $t=50$ (note that we did not plot other correlations for clarity, but they all tend to the same limits). The results are shown in Fig.~\ref{fig:np_g2}, and we see that the correlation tends to the limits that correspond to equal probability for particles, see the explanation and examples in Appendix~\ref{sec:longt_corr}. Correlations between all the other pairs of different sites and same-spin particles have the same limit. For $n=4,6,10$, the fTWA results were compared with Exact Diagonalisation.

\section{fTWA on a large system}

\begin{figure*}
\begin{center}
\includegraphics[height=14 cm, angle =90]{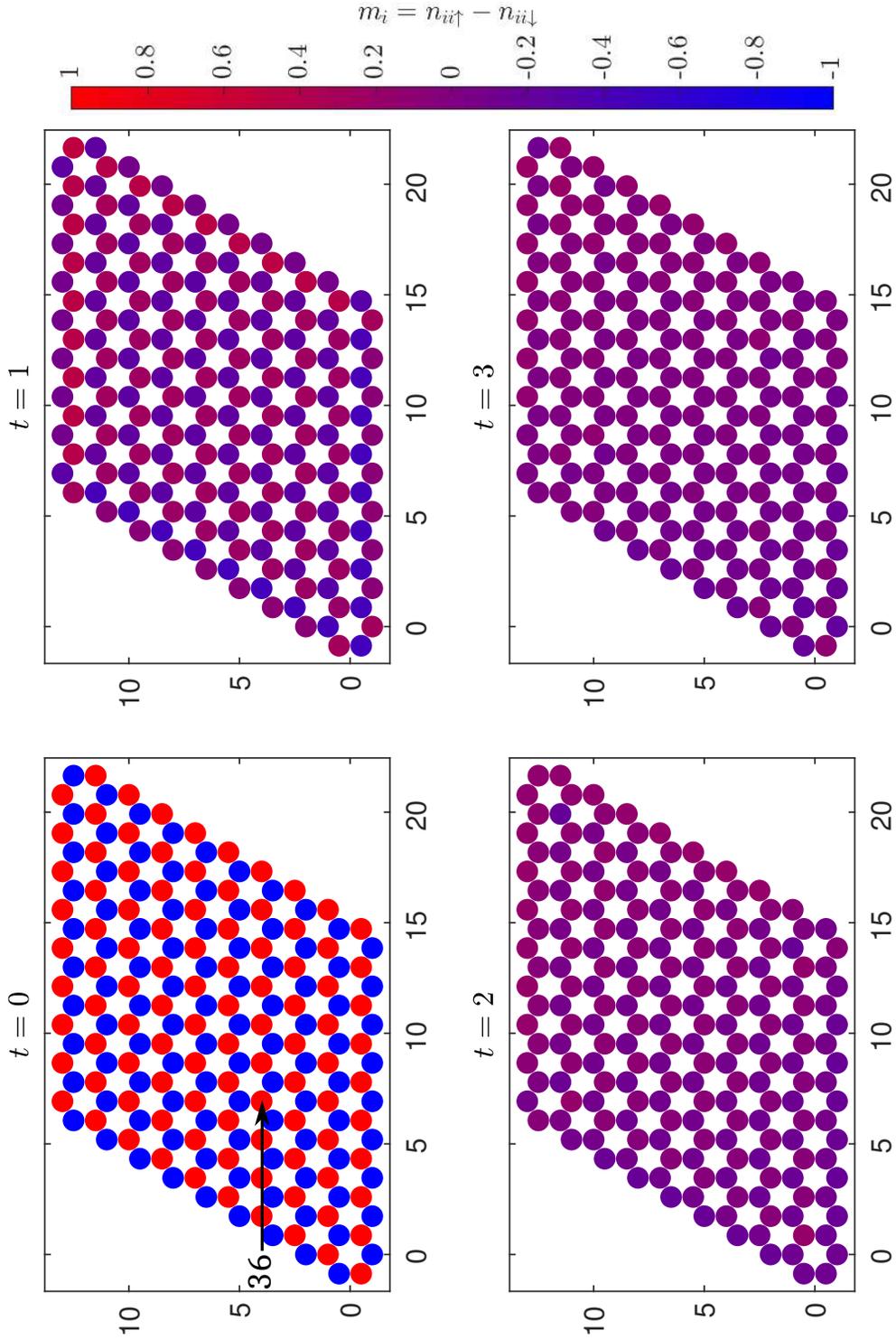}
\caption{(Color online) Four different frames from the dynamics of the magnetic moment $m_i = n_{ii_\uparrow} - n_{ii_\downarrow} $ for times $t=0,1,2,3$. Sites in red have $m_i = 1$, one spin-up particle, sites in blue have $m_i = -1$, one spin-down particle. Sites in purple have $m_i = 0$. For larger times we see that the magnetic moments all stabilize to $m_i\simeq0$, when the system tends to $n_{ii\uparrow} = n_{ii\downarrow} \simeq 0.5$. }
\label{fig:10frames_198sites}
\end{center}
\end{figure*}

Now that we have shown the ability of fTWA to efficiently model small Fermi-Hubbard systems, by a direct comparison to data from Exact Diagonalization, we study a substantially larger system, that is far out of reach for Exact Diagonalisation. We compute the dynamics of fermions described by the Hamiltonian of Eq.~(\ref{eq:FHeq1}) on a graphene-like system with a honeycomb structure involving 198 sites, see the overall geometry in Fig.~\ref{fig:10frames_198sites}. In this figure the value of the magnetisation, $m_i = n_{ii\uparrow} - n_{ii\downarrow}$, is shown for four time frames of the calculation. At $t=0$, the sites are filled with either a spin-up electron or a spin-down electron, represented by red and blue circles in Fig.~\ref{fig:10frames_198sites}, respectively. This represents a starting state similar to that considered in Fig.~\ref{fig:10p_state1}. For the subsequent times, the dynamics is such that the magnetisation at each site approaches zero ($n_{ii\uparrow} \simeq n_{ii\downarrow} \simeq 0.5$), representing an equal occupation of spin-up and spin-down electrons. This is seen most clearly in Fig.~\ref{fig:10frames_198sites} by the purple color of all sites at $t=3$. This result is consistent with experimental data of graphene at equilibrium conditions, that are known to reflect an equal occupation of spin-up and spin-down electrons \cite{katsnelson2007graphene}.

\begin{figure}
\begin{center}
\includegraphics[height=6.5 cm]{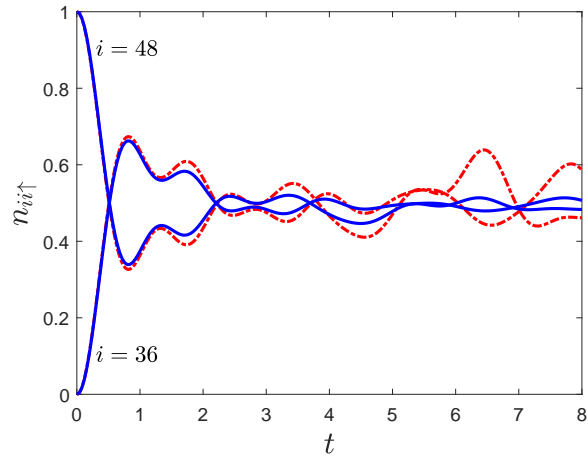}
\caption{(Color online) Dynamics of site occupation for spin-up particles on sites $36$ and $48$ for a 198 atom cluster. Data obtained by HF given by dashed red curves and fTWA full blue curves.} 
\label{fig:198p_niiu}
\end{center}
\end{figure}

For a more accurate view of the dynamics, we have plotted in Fig.~\ref{fig:198p_niiu} the occupation of spin-up electrons on sites $36$ and $48$, $n_{36\uparrow}$ and $n_{48\uparrow}$, i.e. from sites in the middle of the system. Within the short time of the simulation, effects of the boundary of the 198 atom cluster play little role, and the system appears infinite, as reflected in the visible symmetry between $n_{36\uparrow}$ and $n_{48\uparrow}$ in Fig.~\ref{fig:198p_niiu}. This symmetry breaks at later times, around $t\simeq 4$. Note that the data in Fig.~\ref{fig:198p_niiu}  contains results from HF and fTWA calculations and that for short simulation times ($t\leq 3$) we find similar occupations for the two approaches, in contrast to the example in Fig.~\ref{fig:10p_occup} that considered a smaller system. The size of the system gives the site interactions a more predominant role, which we expect makes HF and fTWA better approximations of the real particle dynamics, but clearly ED is out of reach for comparisons. Also, for longer simulation times, the HF method gives highly oscillatory results, as shown in Fig.~\ref{fig:198p_niiu}, which is not the case for the fTWA method.

\begin{figure}
\begin{center}
\includegraphics[height=6.5 cm]{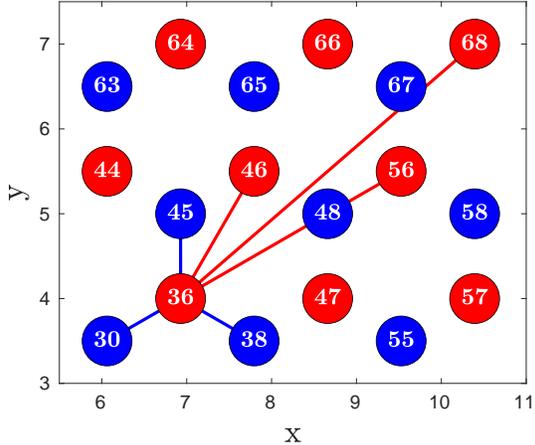}
\caption{(Color online) Zoom in on the 198-sites Fermi-Hubbard system, of Fig.~\ref{fig:10frames_198sites}, at $t=0$. The sites in red starts with a spin-up particle, e.g. $n_{36\uparrow}(0) = 1$, and the blue ones starts with a spin-down particle, e.g. $n_{48\downarrow}(0) = 1$. We will follow the correlation function of spin-up particles between the site $36$ and one of the three closest neighbours (blue lines) in Fig.~\ref{fig:198p_g2_neib}. Then between site $36$ and three distant sites (red lines) in Fig.~\ref{fig:198p_g2_notneib}. } 
\label{fig:198p_g2_pos}
\end{center}
\end{figure}

\begin{figure}
\begin{center}
\includegraphics[height=6.5 cm]{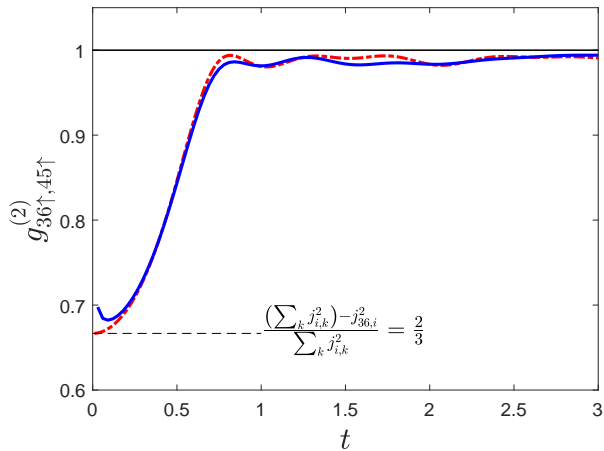}
\caption{(Color online) Dynamics of correlation functions between neighbour sites, here the site $36$ and a neighbour, site $45$. Data obtained by HF given by dashed red curves and fTWA full blue curves. 
} 
\label{fig:198p_g2_neib}
\end{center}
\end{figure}

Similar to the smaller system, we also study for the 198 atom system the correlation functions between different sites. We picked a central site, numbered $36$, to be a representative one  (see the arrow in the $t=0$ subplot in Fig.~\ref{fig:10frames_198sites}), and followed the correlation functions with one of its nearest neighbours as well as with further distant sites, see the geometry in Fig.~\ref{fig:198p_g2_pos}. In Fig.~\ref{fig:198p_g2_neib} we show the correlation functions of spin-up particles with a near neighbour site,  $g^{(2)}_{36\uparrow,45\uparrow}(t)$. The correlation for the three neighbours are initially very similar because of local symmetries in the large system, where edge effects of the cluster play a lesser role. We recognise the short-time limit ($t \rightarrow 0$), where $g^{(2)}(0) = 2/3$, because each site has three neighbours, see the derivation in Appendix~\ref{sec:asymptotes}, and the deviation from the initial value that we saw in the smaller system, in Fig.~\ref{fig:10p_g2_124}. For longer time scales the data in Fig.~\ref{fig:198p_g2_neib} approach a value close to one, that only depends on the total number of sites, see Appendix~\ref{sec:longt_corr}. Again one may note large oscillations with the HF method for longer times.

\begin{figure}
\begin{center}
\includegraphics[height=6.5 cm]{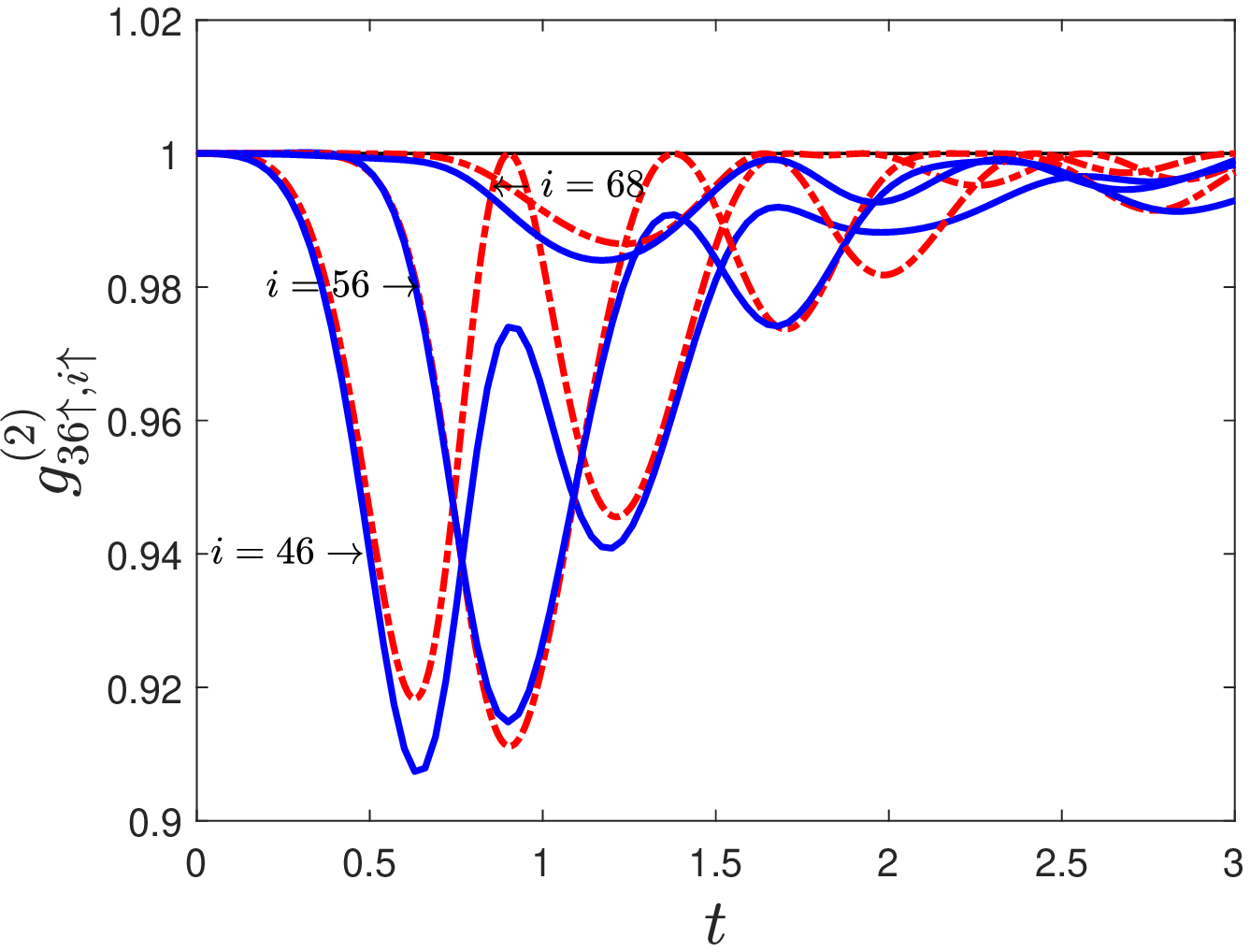}
\caption{(Color online) Dynamics of correlation functions between distant sites, here the site $36$ and sites $46, 56, 68$, see Fig.~\ref{fig:198p_g2_pos}. Data obtained by HF given by dashed red curves and fTWA full blue curves. }
\label{fig:198p_g2_notneib}
\end{center}
\end{figure}

In Fig.~\ref{fig:198p_g2_notneib} we have plotted the correlation functions between the site numbered $36$ and its neighbours at longer distance, see Fig.~\ref{fig:198p_g2_pos} for the geometry. Note that our choice of sites is such that they are initially filled with electrons of the same spin orientation, which means that at $t=0$ the correlation function is one. We can observe from Fig.~\ref{fig:198p_g2_notneib} that the further away two sites are, the longer it takes before $g^{(2)}$ starts to deviate from unity.

\mycomment{
\begin{figure}
\begin{center}
\includegraphics[height=6.5 cm]{198sites_8t_g2_36-45.eps}
\caption{Same but longer } 
\label{fig:198p_8t_g2_pos}
\end{center}
\end{figure} }

\section{Conclusion}

In this work we have studied the quantum dynamics of the electronic structure of graphene-like systems, using an electronic Hamiltonian that allows for hopping and one-site Coulomb repulsion. The analysis is focused on the electron states close to the Fermi level, and are hence limited to $p_z$ orbitals of spin-up or spin-down character centered on each site of a honeycomb lattice site. This allows to study the dynamics of an electronic Hamiltonian that includes the minimum interactions to represent a realistic system, i.e. the electron hopping and on-site Coulomb repulsion. 

We have in this investigation compared three methods with which to solve the time evolution of the electronic system; the Exact Diagonalization technique, the Hartree-Fock (HF) approximation and the fermionic Truncated Wigner Approximation (fTWA). In comparing the three approaches for smaller graphene-like systems we conclude that fTWA reproduces the results of Exact Diagonalization, for significantly longer times compared to HF, and for this reason we have focused on fTWA for larger systems. Previous works of fTWA have focused on long-range interactions \cite{sajna2020semiclassical}, but mean-field and phase-space representation methods have larger difficulties to model on-site interactions because of the more predominant role of quantum effects. Under those conditions, fTWA demonstrates a net improvement over mean-field methods. As shown here, the evolution of site occupations agrees well with exact results, for a period that is three times longer for fTWA compared to HF (see Fig.~\ref{fig:10p_occup}), and its long-time behaviour is also quantitatively recovered. The second-order correlation functions are also found to be well approximated, both on short-time dynamics (Figs.~\ref{fig:10p_g2_124}, \ref{fig:10p_g2_1610}) and long-time dynamics (Fig.~\ref{fig:np_g2}). When comparing large and small systems, the results here are consistent with previous results; that smaller systems exhibit larger fluctuations in e.g. the correlation function, compared to larger ones. 

The improvements of fTWA over mean-field methods come with an acceptable computational cost. A fTWA computation scales as $O(n^2)$ with the number of sites $n$, similar to the HF computation. It needs however multiple repetitions of computations (trajectories) to average upon. This cost is manageable on a single computer for the systems studied here and embarrassingly parallelizable for larger systems.

\section*{Acknowledgments}

We thank Joel Corney and Adam Sajna for valuable discussions. F.~Rousse and M.~\"{O}gren~ are supported by Carl Tryggers foundation and RR-ORU-2021/2022. O.~Eriksson acknowledges support from the Swedish Research Council, the Knut and Alice Wallenberg Foundation, the European Research Council via Synergy Grant 854843 - FASTCORR, Energimyndigheten and eSSENCE.

\bibliography{main_graphene_2022_Magnus.bib} 

\begin{thebibliography}{23}%
\makeatletter
\providecommand \@ifxundefined [1]{%
 \@ifx{#1\undefined}
}%
\providecommand \@ifnum [1]{%
 \ifnum #1\expandafter \@firstoftwo
 \else \expandafter \@secondoftwo
 \fi
}%
\providecommand \@ifx [1]{%
 \ifx #1\expandafter \@firstoftwo
 \else \expandafter \@secondoftwo
 \fi
}%
\providecommand \natexlab [1]{#1}%
\providecommand \enquote  [1]{``#1''}%
\providecommand \bibnamefont  [1]{#1}%
\providecommand \bibfnamefont [1]{#1}%
\providecommand \citenamefont [1]{#1}%
\providecommand \href@noop [0]{\@secondoftwo}%
\providecommand \href [0]{\begingroup \@sanitize@url \@href}%
\providecommand \@href[1]{\@@startlink{#1}\@@href}%
\providecommand \@@href[1]{\endgroup#1\@@endlink}%
\providecommand \@sanitize@url [0]{\catcode `\\12\catcode `\$12\catcode
  `\&12\catcode `\#12\catcode `\^12\catcode `\_12\catcode `\%12\relax}%
\providecommand \@@startlink[1]{}%
\providecommand \@@endlink[0]{}%
\providecommand \url  [0]{\begingroup\@sanitize@url \@url }%
\providecommand \@url [1]{\endgroup\@href {#1}{\urlprefix }}%
\providecommand \urlprefix  [0]{URL }%
\providecommand \Eprint [0]{\href }%
\providecommand \doibase [0]{http://dx.doi.org/}%
\providecommand \selectlanguage [0]{\@gobble}%
\providecommand \bibinfo  [0]{\@secondoftwo}%
\providecommand \bibfield  [0]{\@secondoftwo}%
\providecommand \translation [1]{[#1]}%
\providecommand \BibitemOpen [0]{}%
\providecommand \bibitemStop [0]{}%
\providecommand \bibitemNoStop [0]{.\EOS\space}%
\providecommand \EOS [0]{\spacefactor3000\relax}%
\providecommand \BibitemShut  [1]{\csname bibitem#1\endcsname}%
\let\auto@bib@innerbib\@empty
\bibitem [{\citenamefont {Novoselov}\ \emph {et~al.}(2004)\citenamefont
  {Novoselov}, \citenamefont {Geim}, \citenamefont {Morozov}, \citenamefont
  {Jiang}, \citenamefont {Zhang}, \citenamefont {Dubonos}, \citenamefont
  {Grigorieva},\ and\ \citenamefont {Firsov}}]{novoselov2004electric}%
  \BibitemOpen
  \bibfield  {author} {\bibinfo {author} {\bibfnamefont {K.~S.}\ \bibnamefont
  {Novoselov}}, \bibinfo {author} {\bibfnamefont {A.~K.}\ \bibnamefont {Geim}},
  \bibinfo {author} {\bibfnamefont {S.~V.}\ \bibnamefont {Morozov}}, \bibinfo
  {author} {\bibfnamefont {D.-e.}\ \bibnamefont {Jiang}}, \bibinfo {author}
  {\bibfnamefont {Y.}~\bibnamefont {Zhang}}, \bibinfo {author} {\bibfnamefont
  {S.~V.}\ \bibnamefont {Dubonos}}, \bibinfo {author} {\bibfnamefont {I.~V.}\
  \bibnamefont {Grigorieva}}, \ and\ \bibinfo {author} {\bibfnamefont {A.~A.}\
  \bibnamefont {Firsov}},\ }\href@noop {} {\bibfield  {journal} {\bibinfo
  {journal} {Science}\ }\textbf {\bibinfo {volume} {306}},\ \bibinfo {pages}
  {666} (\bibinfo {year} {2004})}\BibitemShut {NoStop}%
\bibitem [{\citenamefont {Katsnelson}(2007)}]{katsnelson2007graphene}%
  \BibitemOpen
  \bibfield  {author} {\bibinfo {author} {\bibfnamefont {M.~I.}\ \bibnamefont
  {Katsnelson}},\ }\href@noop {} {\bibfield  {journal} {\bibinfo  {journal}
  {Materials Today}\ }\textbf {\bibinfo {volume} {10}},\ \bibinfo {pages} {20}
  (\bibinfo {year} {2007})}\BibitemShut {NoStop}%
\bibitem [{\citenamefont {Geim}\ and\ \citenamefont
  {Novoselov}(2010)}]{geim2010rise}%
  \BibitemOpen
  \bibfield  {author} {\bibinfo {author} {\bibfnamefont {A.~K.}\ \bibnamefont
  {Geim}}\ and\ \bibinfo {author} {\bibfnamefont {K.~S.}\ \bibnamefont
  {Novoselov}},\ }in\ \href@noop {} {\emph {\bibinfo {booktitle} {Nanoscience
  and technology: a collection of reviews from nature journals}}}\ (\bibinfo
  {publisher} {World Scientific},\ \bibinfo {year} {2010})\ pp.\ \bibinfo
  {pages} {11--19}\BibitemShut {NoStop}%
\bibitem [{\citenamefont {Wallace}(1947)}]{wallace1947band}%
  \BibitemOpen
  \bibfield  {author} {\bibinfo {author} {\bibfnamefont {P.~R.}\ \bibnamefont
  {Wallace}},\ }\href@noop {} {\bibfield  {journal} {\bibinfo  {journal}
  {Physical Review}\ }\textbf {\bibinfo {volume} {71}},\ \bibinfo {pages} {622}
  (\bibinfo {year} {1947})}\BibitemShut {NoStop}%
\bibitem [{\citenamefont {McClure}(1956)}]{mcclure1956diamagnetism}%
  \BibitemOpen
  \bibfield  {author} {\bibinfo {author} {\bibfnamefont {J.}~\bibnamefont
  {McClure}},\ }\href@noop {} {\bibfield  {journal} {\bibinfo  {journal}
  {Physical Review}\ }\textbf {\bibinfo {volume} {104}},\ \bibinfo {pages}
  {666} (\bibinfo {year} {1956})}\BibitemShut {NoStop}%
\bibitem [{\citenamefont {Neto}\ \emph {et~al.}(2009)\citenamefont {Neto},
  \citenamefont {Guinea}, \citenamefont {Peres}, \citenamefont {Novoselov},\
  and\ \citenamefont {Geim}}]{neto2009electronic}%
  \BibitemOpen
  \bibfield  {author} {\bibinfo {author} {\bibfnamefont {A.~C.}\ \bibnamefont
  {Neto}}, \bibinfo {author} {\bibfnamefont {F.}~\bibnamefont {Guinea}},
  \bibinfo {author} {\bibfnamefont {N.~M.}\ \bibnamefont {Peres}}, \bibinfo
  {author} {\bibfnamefont {K.~S.}\ \bibnamefont {Novoselov}}, \ and\ \bibinfo
  {author} {\bibfnamefont {A.~K.}\ \bibnamefont {Geim}},\ }\href@noop {}
  {\bibfield  {journal} {\bibinfo  {journal} {Reviews of Modern Physics}\
  }\textbf {\bibinfo {volume} {81}},\ \bibinfo {pages} {109} (\bibinfo {year}
  {2009})}\BibitemShut {NoStop}%
\bibitem [{\citenamefont {Katsnelson}\ \emph {et~al.}(2006)\citenamefont
  {Katsnelson}, \citenamefont {Novoselov},\ and\ \citenamefont
  {Geim}}]{katsnelson2006chiral}%
  \BibitemOpen
  \bibfield  {author} {\bibinfo {author} {\bibfnamefont {M.}~\bibnamefont
  {Katsnelson}}, \bibinfo {author} {\bibfnamefont {K.}~\bibnamefont
  {Novoselov}}, \ and\ \bibinfo {author} {\bibfnamefont {A.}~\bibnamefont
  {Geim}},\ }\href@noop {} {\bibfield  {journal} {\bibinfo  {journal} {Nature
  Physics}\ }\textbf {\bibinfo {volume} {2}},\ \bibinfo {pages} {620} (\bibinfo
  {year} {2006})}\BibitemShut {NoStop}%
\bibitem [{\citenamefont {Katsnelson}(2011)}]{katsnelsonbook}%
  \BibitemOpen
  \bibfield  {author} {\bibinfo {author} {\bibfnamefont {M.~I.}\ \bibnamefont
  {Katsnelson}},\ }in\ \href@noop {} {\emph {\bibinfo {booktitle} {Graphene:
  carbon in two dimensions}}}\ (\bibinfo  {publisher} {Cambridge University
  Press},\ \bibinfo {year} {2011})\ pp.\ \bibinfo {pages} {1--351}\BibitemShut
  {NoStop}%
\bibitem [{\citenamefont {Bolotin}\ \emph {et~al.}(2008)\citenamefont
  {Bolotin}, \citenamefont {Sikes}, \citenamefont {Jiang}, \citenamefont
  {Klima}, \citenamefont {Fudenberg}, \citenamefont {Hone}, \citenamefont
  {Kim},\ and\ \citenamefont {Stormer}}]{bolotin2008ultrahigh}%
  \BibitemOpen
  \bibfield  {author} {\bibinfo {author} {\bibfnamefont {K.~I.}\ \bibnamefont
  {Bolotin}}, \bibinfo {author} {\bibfnamefont {K.~J.}\ \bibnamefont {Sikes}},
  \bibinfo {author} {\bibfnamefont {Z.}~\bibnamefont {Jiang}}, \bibinfo
  {author} {\bibfnamefont {M.}~\bibnamefont {Klima}}, \bibinfo {author}
  {\bibfnamefont {G.}~\bibnamefont {Fudenberg}}, \bibinfo {author}
  {\bibfnamefont {J.}~\bibnamefont {Hone}}, \bibinfo {author} {\bibfnamefont
  {P.}~\bibnamefont {Kim}}, \ and\ \bibinfo {author} {\bibfnamefont {H.~L.}\
  \bibnamefont {Stormer}},\ }\href@noop {} {\bibfield  {journal} {\bibinfo
  {journal} {Solid State Communications}\ }\textbf {\bibinfo {volume} {146}},\
  \bibinfo {pages} {351} (\bibinfo {year} {2008})}\BibitemShut {NoStop}%
\bibitem [{\citenamefont {Panda}\ \emph {et~al.}(2020)\citenamefont {Panda},
  \citenamefont {Ramu}, \citenamefont {Karis}, \citenamefont {Sarkar},\ and\
  \citenamefont {Kamalakar}}]{panda2020ultimate}%
  \BibitemOpen
  \bibfield  {author} {\bibinfo {author} {\bibfnamefont {J.}~\bibnamefont
  {Panda}}, \bibinfo {author} {\bibfnamefont {M.}~\bibnamefont {Ramu}},
  \bibinfo {author} {\bibfnamefont {O.}~\bibnamefont {Karis}}, \bibinfo
  {author} {\bibfnamefont {T.}~\bibnamefont {Sarkar}}, \ and\ \bibinfo {author}
  {\bibfnamefont {M.~V.}\ \bibnamefont {Kamalakar}},\ }\href@noop {} {\bibfield
   {journal} {\bibinfo  {journal} {ACS nano}\ }\textbf {\bibinfo {volume}
  {14}},\ \bibinfo {pages} {12771} (\bibinfo {year} {2020})}\BibitemShut
  {NoStop}%
\bibitem [{\citenamefont {Harrison}(2012)}]{harrison2012electronic}%
  \BibitemOpen
  \bibfield  {author} {\bibinfo {author} {\bibfnamefont {W.~A.}\ \bibnamefont
  {Harrison}},\ }\href@noop {} {\emph {\bibinfo {title} {Electronic structure
  and the properties of solids: the physics of the chemical bond}}}\ (\bibinfo
  {publisher} {Courier Corporation},\ \bibinfo {year} {2012})\BibitemShut
  {NoStop}%
\bibitem [{\citenamefont {Sajna}\ and\ \citenamefont
  {Polkovnikov}(2020)}]{sajna2020semiclassical}%
  \BibitemOpen
  \bibfield  {author} {\bibinfo {author} {\bibfnamefont {A.~S.}\ \bibnamefont
  {Sajna}}\ and\ \bibinfo {author} {\bibfnamefont {A.}~\bibnamefont
  {Polkovnikov}},\ }\href@noop {} {\bibfield  {journal} {\bibinfo  {journal}
  {Physical Review A}\ }\textbf {\bibinfo {volume} {102}},\ \bibinfo {pages}
  {033338} (\bibinfo {year} {2020})}\BibitemShut {NoStop}%
\bibitem [{\citenamefont {Ma}\ \emph {et~al.}(2011)\citenamefont {Ma},
  \citenamefont {Hu}, \citenamefont {Huang},\ and\ \citenamefont
  {Lin}}]{ma2011magnetic}%
  \BibitemOpen
  \bibfield  {author} {\bibinfo {author} {\bibfnamefont {T.}~\bibnamefont
  {Ma}}, \bibinfo {author} {\bibfnamefont {F.}~\bibnamefont {Hu}}, \bibinfo
  {author} {\bibfnamefont {Z.}~\bibnamefont {Huang}}, \ and\ \bibinfo {author}
  {\bibfnamefont {H.-Q.}\ \bibnamefont {Lin}},\ }\href@noop {} {\bibfield
  {journal} {\bibinfo  {journal} {Computer Physics Communications}\ }\textbf
  {\bibinfo {volume} {182}},\ \bibinfo {pages} {52} (\bibinfo {year}
  {2011})}\BibitemShut {NoStop}%
\bibitem [{\citenamefont {Joost}\ \emph {et~al.}(2019)\citenamefont {Joost},
  \citenamefont {Schl{\"u}nzen},\ and\ \citenamefont
  {Bonitz}}]{joost2019femtosecond}%
  \BibitemOpen
  \bibfield  {author} {\bibinfo {author} {\bibfnamefont {J.-P.}\ \bibnamefont
  {Joost}}, \bibinfo {author} {\bibfnamefont {N.}~\bibnamefont
  {Schl{\"u}nzen}}, \ and\ \bibinfo {author} {\bibfnamefont {M.}~\bibnamefont
  {Bonitz}},\ }\href@noop {} {\bibfield  {journal} {\bibinfo  {journal}
  {Physica Status Solidi (b)}\ }\textbf {\bibinfo {volume} {256}},\ \bibinfo
  {pages} {1800498} (\bibinfo {year} {2019})}\BibitemShut {NoStop}%
\bibitem [{\citenamefont {Walls}\ and\ \citenamefont
  {Milburn}(2007)}]{walls2007quantum}%
  \BibitemOpen
  \bibfield  {author} {\bibinfo {author} {\bibfnamefont {D.~F.}\ \bibnamefont
  {Walls}}\ and\ \bibinfo {author} {\bibfnamefont {G.~J.}\ \bibnamefont
  {Milburn}},\ }\href@noop {} {\emph {\bibinfo {title} {Quantum optics}}}\
  (\bibinfo  {publisher} {Springer Science \& Business Media},\ \bibinfo {year}
  {2007})\BibitemShut {NoStop}%
\bibitem [{\citenamefont {Corney}\ and\ \citenamefont
  {Drummond}(2006)}]{corney2006gaussian}%
  \BibitemOpen
  \bibfield  {author} {\bibinfo {author} {\bibfnamefont {J.~F.}\ \bibnamefont
  {Corney}}\ and\ \bibinfo {author} {\bibfnamefont {P.~D.}\ \bibnamefont
  {Drummond}},\ }\href@noop {} {\bibfield  {journal} {\bibinfo  {journal}
  {Physical Review B}\ }\textbf {\bibinfo {volume} {73}},\ \bibinfo {pages}
  {125112} (\bibinfo {year} {2006})}\BibitemShut {NoStop}%
\bibitem [{\citenamefont {{\"O}gren}\ \emph {et~al.}(2010)\citenamefont
  {{\"O}gren}, \citenamefont {Kheruntsyan},\ and\ \citenamefont
  {Corney}}]{ogren2010first}%
  \BibitemOpen
  \bibfield  {author} {\bibinfo {author} {\bibfnamefont {M.}~\bibnamefont
  {{\"O}gren}}, \bibinfo {author} {\bibfnamefont {K.}~\bibnamefont
  {Kheruntsyan}}, \ and\ \bibinfo {author} {\bibfnamefont {J.}~\bibnamefont
  {Corney}},\ }\href@noop {} {\bibfield  {journal} {\bibinfo  {journal} {EPL
  (Europhysics Letters)}\ }\textbf {\bibinfo {volume} {92}},\ \bibinfo {pages}
  {36003} (\bibinfo {year} {2010})}\BibitemShut {NoStop}%
\bibitem [{\citenamefont {Davidson}\ \emph {et~al.}(2017)\citenamefont
  {Davidson}, \citenamefont {Sels},\ and\ \citenamefont
  {Polkovnikov}}]{davidson2017semiclassical}%
  \BibitemOpen
  \bibfield  {author} {\bibinfo {author} {\bibfnamefont {S.~M.}\ \bibnamefont
  {Davidson}}, \bibinfo {author} {\bibfnamefont {D.}~\bibnamefont {Sels}}, \
  and\ \bibinfo {author} {\bibfnamefont {A.}~\bibnamefont {Polkovnikov}},\
  }\href@noop {} {\bibfield  {journal} {\bibinfo  {journal} {Annals of
  Physics}\ }\textbf {\bibinfo {volume} {384}},\ \bibinfo {pages} {128}
  (\bibinfo {year} {2017})}\BibitemShut {NoStop}%
\bibitem [{\citenamefont {Sidje}(1998)}]{sidje1998expokit}%
  \BibitemOpen
  \bibfield  {author} {\bibinfo {author} {\bibfnamefont {R.~B.}\ \bibnamefont
  {Sidje}},\ }\href@noop {} {\bibfield  {journal} {\bibinfo  {journal} {ACM
  Transactions on Mathematical Software (TOMS)}\ }\textbf {\bibinfo {volume}
  {24}},\ \bibinfo {pages} {130} (\bibinfo {year} {1998})}\BibitemShut
  {NoStop}%
\bibitem [{\citenamefont {Rahav}\ and\ \citenamefont
  {Mukamel}(2009)}]{rahav2009gaussian}%
  \BibitemOpen
  \bibfield  {author} {\bibinfo {author} {\bibfnamefont {S.}~\bibnamefont
  {Rahav}}\ and\ \bibinfo {author} {\bibfnamefont {S.}~\bibnamefont
  {Mukamel}},\ }\href@noop {} {\bibfield  {journal} {\bibinfo  {journal}
  {Physical Review B}\ }\textbf {\bibinfo {volume} {79}},\ \bibinfo {pages}
  {165103} (\bibinfo {year} {2009})}\BibitemShut {NoStop}%
\bibitem [{\citenamefont {Sinatra}\ \emph {et~al.}(2002)\citenamefont
  {Sinatra}, \citenamefont {Lobo},\ and\ \citenamefont
  {Castin}}]{sinatra2002truncated}%
  \BibitemOpen
  \bibfield  {author} {\bibinfo {author} {\bibfnamefont {A.}~\bibnamefont
  {Sinatra}}, \bibinfo {author} {\bibfnamefont {C.}~\bibnamefont {Lobo}}, \
  and\ \bibinfo {author} {\bibfnamefont {Y.}~\bibnamefont {Castin}},\
  }\href@noop {} {\bibfield  {journal} {\bibinfo  {journal} {Journal of Physics
  B: Atomic, Molecular and Optical Physics}\ }\textbf {\bibinfo {volume}
  {35}},\ \bibinfo {pages} {3599} (\bibinfo {year} {2002})}\BibitemShut
  {NoStop}%
\bibitem [{\citenamefont {Wick}(1950)}]{wick1950evaluation}%
  \BibitemOpen
  \bibfield  {author} {\bibinfo {author} {\bibfnamefont {G.-C.}\ \bibnamefont
  {Wick}},\ }\href@noop {} {\bibfield  {journal} {\bibinfo  {journal} {Physical
  Review}\ }\textbf {\bibinfo {volume} {80}},\ \bibinfo {pages} {268} (\bibinfo
  {year} {1950})}\BibitemShut {NoStop}%
\bibitem [{\citenamefont {Davidson}\ and\ \citenamefont
  {Polkovnikov}(2015)}]{davidson2015s}%
  \BibitemOpen
  \bibfield  {author} {\bibinfo {author} {\bibfnamefont {S.~M.}\ \bibnamefont
  {Davidson}}\ and\ \bibinfo {author} {\bibfnamefont {A.}~\bibnamefont
  {Polkovnikov}},\ }\href@noop {} {\bibfield  {journal} {\bibinfo  {journal}
  {Physical Review Letters}\ }\textbf {\bibinfo {volume} {114}},\ \bibinfo
  {pages} {045701} (\bibinfo {year} {2015})}\BibitemShut {NoStop}%
\end{thebibliography}%
\bibliographystyle{apsrev4-1}

\newpage

\appendix

{\centerline {\huge Appendix}}

\section{Derivation of differential equations for the quantum dynamics}
\label{sec:Motion_eq}

In this appendix, we outline some of the details needed for the computations described in the main text. In the following, the indices written as Greek letters are pairs of position and spin, $\alpha = (a, \sigma_a)$.

\subsection{Heisenberg equation of motion}

In the Heisenberg picture, the operators are time dependent and evolve according to the differential equation:
\begin{equation}
    \frac{d}{d t} \hat{A}  = i \left[\hat{H}, \hat{A} \right] \text{.}
\end{equation}
Hence for a Fermi-Hubbard Hamiltonian described in Eq.~(\ref{eq:FHeq1}) with symmetric electron-electron interactions $u_{\alpha \beta} = u_{\beta \alpha }$, the differential equation of a bilinear operators like $\hat{c}_{\alpha}^{\dagger} \hat{c}_{\beta}$ is

\begin{equation}
\label{eq:Heis_1}
\begin{split}
    \frac{\partial}{\partial t} \hat{c}^{\dagger}_{\alpha} \hat{c}_{\beta}  &= i \sum_{\mu} \bigg( (j_{\beta \mu} \hat{c}_{\alpha}^{\dagger} \hat{c}_{ \mu}  -  j_{\mu \alpha} \hat{c}_{\mu}^{\dagger} \hat{c}_{\beta} ) \\
    &+   (u_{\mu \alpha}  - u_{\beta \mu}) \hat{c}^{\dagger}_{\alpha} \hat{c}_{\mu}^{\dagger} \hat{c}_{ \mu}  \hat{c}_{\beta}   \bigg) \text{.}
\end{split}
\end{equation}
This differential equation for the number operators cannot be solved directly in practice, and we need to invoke approximations to obtain differential equations for complex numbers. 

\subsection{Mean-field approximation}
\label{sec:Motion_eq_hf}

In the Hartree-Fock approximation, the many-body wave-function is described as a product of one-body wave-functions, hence the system is always a single Slater determinant. That allows us to reduce the description of the electrons to a  single-particle density matrix with
\begin{equation}
\begin{split}
      n_{\alpha \beta} \equiv \langle \hat{c}_{\alpha}^{\dagger} \hat{c}_{\beta} \rangle \text{.}
     \end{split}
\end{equation}    
For the two-body operators, we use Wick's theorem \cite{rahav2009gaussian, wick1950evaluation}: 
 
\begin{equation}
\begin{split}
       \langle \hat{c}_{\alpha}^{\dagger} \hat{c}_{\beta}^{\dagger} \hat{c}_{\gamma} \hat{c}_{\delta}  \rangle \equiv n_{\alpha \delta} n_{\beta \gamma} - n_{\alpha \gamma} n_{\beta \delta} \text{.}
     \end{split}
     \label{eq:wickthm}
\end{equation}    
Which gives the differential equation for the single-particle density matrix: 

\begin{equation}
\label{eq:Heis_HF1}
\begin{split}
    \frac{\partial}{\partial t} n_{\alpha \beta}  &= i \sum_{\mu} \bigg( (j_{\beta \mu} n_{\alpha \mu}  -  j_{\mu \alpha} n_{\mu \beta} ) \\
    &+    (u_{\mu \alpha}  - u_{\beta \mu}) ( n_{\alpha \beta} n_{\mu \mu} - n_{\alpha \mu} n_{\mu \beta} ) \bigg) \text{.}
\end{split}
\end{equation}

In the Hamiltonian~(\ref{eq:FHeq1}), no term allows the particle's spin to flip, ($j_{\alpha \beta} = j_{ab}\delta_{\sigma_a \sigma_b}$) so variables representing spin-flip are always null and are neglected. Moreover, the electron-electron interaction is only between opposite-spin particles, $u_{\alpha \beta} = u_{ab} \delta_{\sigma_a \bar{\sigma_b}}$. We denote $n_{ij\sigma} = n_{i\sigma,j\sigma}$, and Eq.~(\ref{eq:Heis_HF1}) then becomes

\begin{equation}
\label{eq:Heis_HF2}
\begin{split}
       \frac{\partial }{\partial t} n_{ij \sigma} &=  i \sum_{ l} \big( j_{j l}    n_{i l \sigma}  -   j_{l i} n_{lj\sigma} +  (u_{il} -u_{l j}) n_{ll \bar{\sigma}}  n_{i j \sigma} \big) \text{.}\\
     \end{split}
\end{equation}    
These are the Hartree-Fock differential equations, as presented in Ref.~\cite{rahav2009gaussian}.

\subsection{Fermionic Truncated Wigner approximation}
\label{sec:Motion_eq_ftwa}

The fTWA is one of the phase-space representation methods, where we use a distribution to describe the electrons density matrix and map operators averages to the distribution's moments. For the fTWA in systems with constant number of particles, we can choose the symmetrically ordered one-body operator $\hat{E}^{\alpha}_{\beta}$ we defined in Eq.~(\ref{eq:ps_var}) and map it to the phase-space variables $\rho_{\alpha \beta}$. First-order and second-order moments of the phase-space variable distribution are linked respectively to one-body and two-body operators:

\begin{equation}
\begin{split}
    \overline{\rho_{\alpha \beta}} = \int W(\rho) \rho_{\alpha \beta} d\rho &= \langle \hat{E}_{\alpha}^{\beta} \rangle \text{,}\\
    \overline{\rho_{\alpha \beta}^*\rho_{\mu \nu}} = \int W(\rho) \rho^*_{\alpha \beta} \rho_{\mu \nu} d\rho &= \frac{1}{2}\langle \hat{E}_{\beta}^{\alpha} \hat{E}_{\mu}^{\nu} + \hat{E}_{\mu}^{\nu}\hat{E}_{\beta}^{\alpha}\rangle \text{.} \\
\end{split}
\label{eq:ftwa_rel}
\end{equation}
From those equations we can find the links between first- and second-order stochastic averages of phase-space variables and quantum operators in e.g. Eq.~(\ref{eq:fTWA_moments}).

To obtain the equations of motion of the phase-space distribution, we follow the work of Polkovnikov~\cite{davidson2015s, davidson2017semiclassical}. 
Here a Jordan-Schwinger mapping from the fermionic operators $\hat{E}$ to pair-bosonic operators was introduced, then  the bosonic Truncated Wigner Approximation formalism can be applied. It was shown that the equation of motion for the phase-space variables $\rho_{\alpha \beta}$ are determined by the Poisson brackets

\begin{equation}
\begin{split}
       \frac{\partial }{\partial t} \rho_{\alpha \beta} &=  i\{ \rho_{\alpha \beta}, H_W \} = i\sum_{\gamma \delta} f(\alpha, \beta, \mu, \nu, \gamma, \delta) \frac{\partial H_W}{\partial \rho_{\mu \nu}}\rho_{\gamma \delta} \text{,}
       \label{eq:ap_fTWAeq_Pol}
     \end{split}
\end{equation} 
with $f$ being the structure constants of the bilinear operators;

\begin{equation}
\begin{split}[\hat{E}_{\beta}^{\alpha}, \hat{E}_{\nu}^{\mu}] &= \sum_{\gamma \delta} f(\alpha, \beta, \mu, \nu, \gamma, \delta) \hat{E}_{\delta}^{\gamma}\\
&= \delta_{\mu \beta}  \hat{E}_{\nu}^{\alpha} -  \delta_{\alpha \nu}  \hat{E}_{\beta}^{\mu} \text{.}
     \end{split}
\end{equation}    
In the expression above, $H_W$ is the Hamiltonian in the $\rho_{\alpha \beta}$ variables: 

\begin{equation}
   H_W = -  \sum_{  i,j , \sigma} j_{ij} \rho_{ij \sigma} +  \sum_{i \geq j} u_{ij}  \left( \rho_{ii \uparrow} + \frac{1}{2}\ \right)\left( \rho_{jj \downarrow} + \frac{1}{2}\ \right) \text{,}
       \label{eq:ftwa_h}
\end{equation}
which gives us the differential equation~(\ref{eq:drho}).

We may not be able to reproduce all exact initial Wigner function with all high-order moments, but within the accuracy of the truncation, the two first moments are sufficient. So we approximate $W(\rho,0)$ with a gaussian distribution having the same first- and second-order moments (mean and covariance). From the relations~(\ref{eq:ftwa_rel}) we compute the covariance between phase-space variables:

\begin{equation}
\begin{split}
    \text{cov}(\rho^*_{\alpha \beta}, \rho_{\mu \nu} ) &= \overline{\rho^*_{\alpha \beta} \rho_{\mu \nu}} - \overline{\rho^*_{\alpha \beta}}  \overline{ \rho_{\mu \nu}}\\
    &= \frac{1}{2} \langle \hat{c}^{\dagger}_{\beta} \hat{c}_{\alpha} \hat{c}^{\dagger}_{\mu} \hat{c}_{\nu} + \hat{c}^{\dagger}_{\mu} \hat{c}_{\nu} \hat{c}^{\dagger}_{\beta}   \hat{c}_{\alpha}\rangle - \langle \hat{c}^{\dagger}_{\beta} \hat{c}_{\alpha} \rangle \langle\hat{c}^{\dagger}_{\mu} \hat{c}_{\nu}\rangle\\
    &= \frac{1}{2}(\tilde{n}_{\mu \alpha} n_{\beta \nu} + \tilde{n}_{\beta \nu} n_{\mu \alpha}) \text{,}
\end{split}
\end{equation}
with $\tilde{n}_{\alpha \beta} = \delta_{\alpha \beta} - n_{ \alpha \beta}$.
In a diagonalising basis, $n_{ \alpha \beta} = 0$ if $\alpha \neq \beta$, so the only non zero terms are  
\begin{equation}
\begin{split}
    \text{cov}(\rho^*_{\alpha \beta}, \rho_{\alpha \beta} ) &= \frac{1}{2}(\tilde{n}_{\alpha \alpha} n_{\beta \beta} + \tilde{n}_{\beta \beta} n_{\alpha \alpha}) \\
     &= \frac{1}{2}(n_{\beta \beta} + n_{\alpha \alpha} - 2n_{\beta \beta}n_{\alpha \alpha} ) \text{.}\\
\end{split}
\end{equation}
Hence we arrive at Eq.~(\ref{eq:ic_fTWA}) for the random starting point of trajectories.

\section{Exact Diagonalisation}

\subsection{Exact Diagonalisation basis}
\label{sec:edb}

In Eq.~(\ref{eq:shro}) describing the wave-function evolution, $\ket{\Psi(t)}$ is written in a finite basis $\ket{\psi_i}$ of Slater determinants of $p$ particles:

\begin{align}
    \ket{\Psi(t)} &= \sum_i b_i(t) \ket{\psi_i} \text{ , }\\
    \ket{\psi_i} &=  \hat{c}_{\alpha_{1,i}}^{\dagger} \hat{c}_{\alpha_{2,i}}^{\dagger} ... \hat{c}_{\alpha_{p,i}}^{\dagger} \ket{0} \text{ .}
\end{align}
If our system has $n$ sites and $p$ particles, we need $C^{p}_{n} = \frac{n!}{p!(n-p)!}$ basis functions, which explains the impossibility to model systems of more than $20$ sites on a standard computer.

\subsection{Analysis of the ground state}
\label{sec:gs}

\begin{figure}
\begin{center}
\includegraphics[height=6.5 cm]{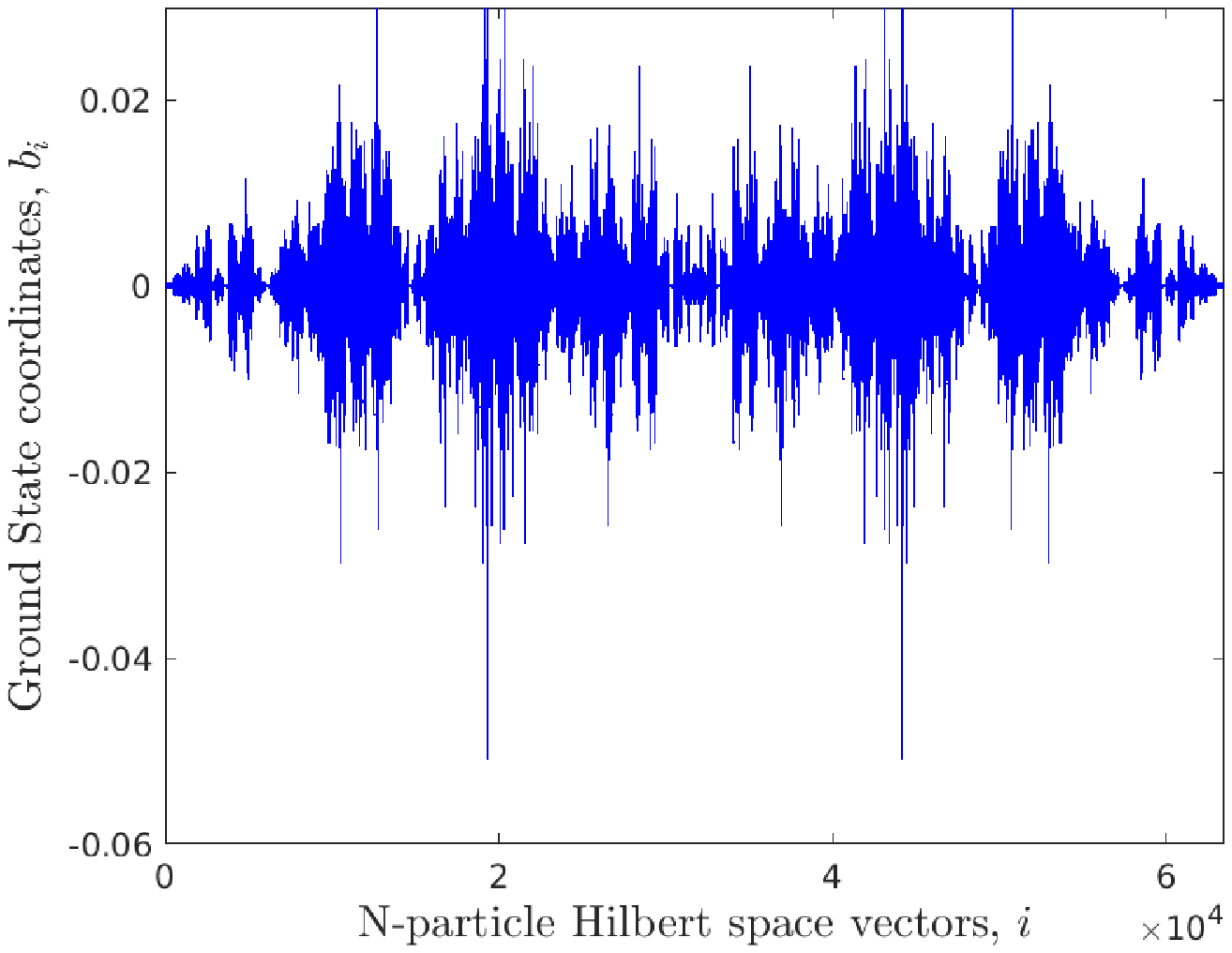}
\caption{Ground state representation of the Fermi-Hubbard system with 10 sites, 5 spin-up and 5 spin-down particles. If $ \ket{\Psi_{gs}}  = \sum_i b_i \ket{\psi_i} $, the x-axis represents the N-particle Hilbert space vector indices $i$ in a given order, and the y-axis represents the value of the coefficients of those basis vectors, $b_i$. We have chosen the first basis vector with the larger absolute coordinate, $b_i \simeq -0.05$, as initial condition for the dynamics reported in the main text. }
\label{fig:10p_gs}
\end{center}
\end{figure}

As initial condition for the dynamics, we are looking for a position eigenstate that has the largest overlap with the ground state of the Hamiltonian~(\ref{eq:FHeq1}). In Fig.~\ref{fig:10p_gs}, we have plotted a representation of the ground state $\ket{\Psi_{gs}} = \sum_{i} b_i \ket{\psi_i}$. On the x-axis are the indices of the N-particle Hilbert basis vectors and on the y-axis are the coordinates of the ground state in this basis, $b_i$. We find two vectors that stand out with $b_i \simeq -0.05$, the first one is the initial condition we have chosen, see Fig.~\ref{fig:10p_state1}, the other one is its spin symmetric state.

When we let the dynamics evolve, we have found that the density on all sites tends to $0.5$ for large times. The corresponding $t\rightarrow \infty$ wave-function is a broad distribution of all N-particle Hilbert space vectors, $\ket{\Psi(t)}\rightarrow \sum_i b_i(\infty) \ket{\psi_i}$ with $|b_i(\infty)|^2 \simeq 1/N$.

\section{Derivation of asymptotes of correlation functions}
\label{sec:asymptotes}

\subsection{Short-time limit of correlations}

We here show that the fTWA equations and the initial conditions leads to the correct values of the  correlation functions for neighbour sites in the short-time limit. Lets compute $g^{(2)}_{1\uparrow, 2\uparrow}(0)$, i.e. the correlation of spin-up particles between site $1$ and site $2$ in a system like in Fig.~\ref{fig:10p_state1} when $t \rightarrow 0$. From Eq.~(\ref{eq:g2_fTWA}) we have,

\mycomment{
\begin{equation}
    g^{(2)}_{1\uparrow, 2\uparrow } = \frac{\overline{\rho_{11\uparrow} \rho_{22\uparrow}} + (\overline{ \rho_{11\uparrow}} + \overline{ \rho_{22\uparrow} }  ) / 2+ 1/4} {(\overline{ \rho_{11\uparrow}}  + 1 / 2) (\overline{ \rho_{22\uparrow}}  + 1/ 2 )} \text{  .}
     \label{eq:g2_all}
\end{equation}
}

\begin{equation}
    g^{(2)}_{1\uparrow, 2\uparrow } = \frac{\langle (\rho_{11\uparrow}  + 1 / 2 )( \rho_{22\uparrow}  + 1/ 2 ) \rangle} {\langle \rho_{11\uparrow}  + 1 / 2 \rangle \langle \rho_{22\uparrow}  + 1/ 2 \rangle} \text{  .}
     \label{eq:g2_all}
\end{equation}

In this appendix, the brackets $\langle ... \rangle$ denote stochastic average. From the initial conditions, we can first derive the first order approximation of the off-diagonal terms $\rho_{12\uparrow}$, $\rho_{21\uparrow}$, $\rho_{23\uparrow}$ and $\rho_{32\uparrow}$:

\begin{align*}
       \frac{\partial }{\partial t} \rho_{12\uparrow} &=  i \sum_{ k}  ( j_{2 k}   \rho_{1 k \uparrow}  -   j_{k 1} \rho_{k2\uparrow} ) + i u \rho_{12 \uparrow} (\rho_{11 \downarrow}-\rho_{22 \downarrow})  \\
        &=  i \big(j_{12} (\rho_{11\uparrow} - \rho_{22\uparrow} ) + j_{32} \rho_{13\uparrow} - j_{14} \rho_{42\uparrow}  \big)\\
        & ~~+ i u \rho_{12 \uparrow} (\rho_{11 \downarrow}-\rho_{22 \downarrow}) \text{  .}
\end{align*}

Now, since at $t=0$, $n_{11\uparrow}(0) = n_{33\uparrow}(0) = 1$, $n_{22\uparrow}(0) = n_{44\uparrow}(0) = 0$, and $ \rho_{13\uparrow}(0) = \rho_{42\uparrow}(0) = 0$, see the initial conditions~(\ref{eq:ic_fTWA}) for the system in Fig.~\ref{fig:10p_state1}, we have:
\begin{align*}
\frac{d}{dt} \rho_{12\uparrow} (0) &= - i u \frac{\xi_{12\uparrow}}{\sqrt{2}} + i j_{12}.\\
\end{align*}
Hence, for the short-time dynamics, i.e. for $t = \epsilon \ll 1$, 
\begin{align*}
\rho_{12\uparrow}(\epsilon) &= \frac{\xi_{12\uparrow}}{\sqrt{2}} + \int_0^{\epsilon} \frac{d}{dt} \rho_{12\uparrow}  dt \\
&= \frac{\xi_{12\uparrow}}{\sqrt{2}} + \int_0^{\epsilon} \left( \frac{d}{dt} \rho_{12\uparrow} \bigg|_{t=0} + O(t) \right) dt \\
&= \frac{\xi_{12\uparrow}}{\sqrt{2}} + i j_{12} \epsilon - i u \frac{\xi_{12\uparrow}}{\sqrt{2}}  \epsilon + O(\epsilon^2) \text{  .} 
\label{eq:ap_g2_12}
\end{align*}
In the same way, we obtain:

\begin{equation}
\begin{split}
 \rho_{21\uparrow} (\epsilon) &= \frac{\xi_{21\uparrow}}{\sqrt{2}} - i j_{12} \epsilon + i u \frac{\xi_{21\uparrow}}{\sqrt{2}}  \epsilon + O(\epsilon^2),\\
\rho_{23\uparrow} (\epsilon) &= \frac{\xi_{23\uparrow}}{\sqrt{2}}  - i j_{32} \epsilon + i u \frac{\xi_{23\uparrow}}{\sqrt{2}} \epsilon + O(\epsilon^2) , \\
 \rho_{32\uparrow} (\epsilon) &= \frac{\xi_{32\uparrow}}{\sqrt{2}} + i j_{32} \epsilon - i  u \frac{\xi_{32\uparrow}}{\sqrt{2}} \epsilon + O(\epsilon^2) .  
 \end{split}
 \label{eq:ap_offDiagTerms}
\end{equation}

With these expressions, we can derive approximations of $\rho_{11\uparrow}$ and $\rho_{22\uparrow}$ with a second-order accuracy in $t$. For arbitrary $t$, we have 

\begin{align*}
\frac{d}{dt} \rho_{11\uparrow} &=  i \sum_{ k} (j_{2 k}    \rho_{1 k \uparrow}  -   j_{k 1} \rho_{k1\uparrow} )+ i u \rho_{11 \uparrow} (\rho_{11 \downarrow}-\rho_{11 \downarrow})   \\
&=   i j_{12}  (\rho_{12\uparrow}-\rho_{21\uparrow}) + i j_{14} ( \rho_{14\uparrow}-\rho_{41\uparrow} ) ) \text{  .}
\end{align*}
Hence, for $t=\epsilon \ll 1$, we can insert the off-diagonal terms of Eq.~(\ref{eq:ap_offDiagTerms}):

\begin{scriptsize}
\begin{align*}
& i \frac{d}{dt} \rho_{11\uparrow}(\epsilon)  \\ 
=&  j_{12} \left( \frac{\xi^*_{12\uparrow}}{\sqrt{2}} - i j_{12} \epsilon + i u  \frac{\xi^*_{12\uparrow}}{\sqrt{2}}  \epsilon -  \frac{\xi_{12\uparrow}}{\sqrt{2}} - i j_{12} \epsilon + i u  \frac{\xi_{12\uparrow}}{\sqrt{2}}   \epsilon  \right)  \\
 &+ j_{14} \left( \frac{\xi^*_{14\uparrow}}{\sqrt{2}} - i j_{14} \epsilon + i u \frac{\xi^*_{14\uparrow}}{\sqrt{2}}   \epsilon - \frac{\xi_{14\uparrow}}{\sqrt{2}} - i j_{14} \epsilon + i u \frac{\xi_{14\uparrow}}{\sqrt{2}} \epsilon  \right) \\
 &+ O(\epsilon^2) \\
 =& i j_{12}  \frac{\xi_{12\uparrow} -\xi^*_{12\uparrow}}{\sqrt{2}} - 2 j_{12}^2 \epsilon  +  u j_{12} \frac{\xi_{12\uparrow} + \xi^*_{12\uparrow}}{\sqrt{2}} \epsilon  \\
 &+ i j_{14} \frac{\xi_{14\uparrow} -\xi^*_{14\uparrow}}{\sqrt{2}} - 2 j_{14}^2 \epsilon  +  u j_{14} \frac{\xi_{14\uparrow} + \xi^*_{14\uparrow}}{\sqrt{2}} \epsilon + O(\epsilon^2) \\
=& - j_{12}  \eta^{(2)}_{12\uparrow} - 2 j_{12}^2 \epsilon  +  u j_{12} \eta^{(1)}_{12\uparrow} \epsilon  - j_{14}  \eta^{(2)}_{14\uparrow} - 2 j_{14}^2 \epsilon  \\
&+  u j_{14} \eta^{(1)}_{14\uparrow} \epsilon + O(\epsilon^2) \\
=& - j_{12}  \eta^{(2)}_{12\uparrow} - j_{14}  \eta^{(2)}_{14\uparrow} + ( - 2 j_{12}^2   - 2 j_{14}^2  +  u j_{12} \eta^{(1)}_{12\uparrow} \\
&+  u j_{14} \eta^{(1)}_{14\uparrow} ) \epsilon + O(\epsilon^2) \text{  ,}
\end{align*}
\end{scriptsize}

such that:

\begin{equation}
\begin{split}
\rho_{11\uparrow}(\epsilon) &= \frac{1}{2} + \int_0^{\epsilon} \frac{d }{dt} \rho_{11\uparrow} dt \\
 &= \frac{1}{2}  - ( j_{12} \eta^{(2)}_{12\uparrow} +  j_{14} \eta^{(2)}_{14\uparrow}) \epsilon  - ( j_{12}^2 +  j_{14}^2 ) \epsilon^2 \\
 &+  \frac{u}{2}(j_{12} \eta^{(1)}_{12\uparrow} +  j_{14} \eta^{(1)}_{14\uparrow} ) \epsilon^2 + O(\epsilon^3) \text{ .}
 \end{split}
 \label{eq:ro11u}
\end{equation}
and in the same way:

\begin{equation}
\begin{split}
\rho_{22\uparrow}(\epsilon) &= - \frac{1}{2}  + (j_{12}  \eta^{(2)}_{12\uparrow} - j_{32}  \eta^{(2)}_{23\uparrow}) \epsilon +  ( j_{12}^2 + j_{32}^2 ) \epsilon^2 \\ & -   \frac{u}{2} ( j_{12}\eta^{(1)}_{12\uparrow} +  j_{32} \eta^{(1)}_{23\uparrow} ) \epsilon^2 + O(\epsilon^3) .
 \label{eq:ro22u}
  \end{split}
\end{equation}

We now obtain from Eqs.~(\ref{eq:ro11u}) and (\ref{eq:ro22u}) first the denominator of $g^{(2)}_{1\uparrow , 2\uparrow}$:

\mycomment{
\begin{align*}
&\langle \rho_{11\uparrow}+\frac{1}{2}\rangle\langle  \rho_{22\uparrow}+\frac{1}{2} \rangle \\
    = & \langle 1 - ( j_{12} \eta^{(2)}_{12\uparrow} +  j_{14} \eta^{(2)}_{14\uparrow}) \epsilon  - ( j_{12}^2 +  j_{14}^2 ) \epsilon^2 \\
    & ~ + u \frac{1}{2}(j_{12} \eta^{(1)}_{12\uparrow} +  j_{14} \eta^{(1)}_{14\uparrow} ) \epsilon^2 \rangle \\
    & ~ \langle   (j_{12}  \eta^{(2)}_{12\uparrow} - j_{32}  \eta^{(2)}_{23\uparrow}) \epsilon +   ( j_{12}^2 + j_{32}^2 ) \epsilon^2 \\
    & ~ - u  \frac{1}{2} ( j_{12}\eta^{(1)}_{12\uparrow} +  j_{32} \eta^{(1)}_{23\uparrow} ) \epsilon^2  \rangle + O(\epsilon^3) \\
    = & \langle   j_{12} \eta^{(2)}_{12}  - j_{23}  \eta^{(2)}_{23\uparrow} \rangle \epsilon  + ( j_{12}^2 + j_{23}^2 ) \epsilon^2  - u  \frac{1}{2} \langle j_{12}\eta^{(1)}_{12\uparrow} +  j_{32} \eta^{(1)}_{23\uparrow} \rangle  \epsilon^2 \\
    & ~ +( - 2 j_{12}^2 \langle \eta^{(2)}_{12\uparrow}\rangle^2 + 2  j_{12} j_{23}\langle \eta^{(2)}_{12\uparrow}\rangle \langle \eta^{(2)}_{23\uparrow} \rangle- j_{14}j_{12}\langle \eta^{(2)}_{14\uparrow}\rangle \langle \eta^{(2)}_{12\uparrow} \rangle \\
    & ~ + j_{14}j_{23}\langle \eta^{(2)}_{14\uparrow}\rangle \langle \eta^{(2)}_{23\uparrow} \rangle )\epsilon^2 + O(\epsilon^3) \text{  .}\\
\end{align*}  
}

\begin{scriptsize}

\begin{align*}
&\langle \rho_{11\uparrow}+\frac{1}{2}\rangle\langle  \rho_{22\uparrow}+\frac{1}{2} \rangle \\
    = & \langle 1 - ( j_{12} \eta^{(2)}_{12\uparrow} +  j_{14} \eta^{(2)}_{14\uparrow}) \epsilon  - ( j_{12}^2 +  j_{14}^2 ) \epsilon^2  +  \frac{u}{2}(j_{12} \eta^{(1)}_{12\uparrow} +  j_{14} \eta^{(1)}_{14\uparrow} ) \epsilon^2 \rangle \\
    & ~ \langle   (j_{12}  \eta^{(2)}_{12\uparrow} - j_{32}  \eta^{(2)}_{23\uparrow}) \epsilon +   ( j_{12}^2 + j_{32}^2 ) \epsilon^2 -   \frac{u}{2} ( j_{12}\eta^{(1)}_{12\uparrow} +  j_{32} \eta^{(1)}_{23\uparrow} ) \epsilon^2  \rangle \\
    & ~ + O(\epsilon^3) \\
    = & \langle   j_{12} \eta^{(2)}_{12\uparrow}  - j_{23}  \eta^{(2)}_{23\uparrow} \rangle \epsilon  + ( j_{12}^2 + j_{23}^2 ) \epsilon^2  -  \frac{u}{2} \langle j_{12}\eta^{(1)}_{12\uparrow} +  j_{32} \eta^{(1)}_{23\uparrow} \rangle  \epsilon^2 \\
    & ~ +( - 2 j_{12}^2 \langle \eta^{(2)}_{12\uparrow}\rangle^2 + 2  j_{12} j_{23}\langle \eta^{(2)}_{12\uparrow}\rangle \langle \eta^{(2)}_{23\uparrow} \rangle- j_{14}j_{12}\langle \eta^{(2)}_{14\uparrow}\rangle \langle \eta^{(2)}_{12\uparrow} \rangle \\
    & ~ + j_{14}j_{23}\langle \eta^{(2)}_{14\uparrow}\rangle \langle \eta^{(2)}_{23\uparrow} \rangle )\epsilon^2 + O(\epsilon^3) \text{  .}\\
\end{align*}   
\end{scriptsize}

\mycomment{
\begin{scriptsize}
\begin{align*}
&( \overline{\rho_{11\uparrow}}+\frac{1}{2})(  \overline{\rho_{22\uparrow}}+\frac{1}{2} ) \\
    = & ( 1 - ( j_{12} \overline{\eta^{(2)}_{12\uparrow}} +  j_{14} \overline{\eta^{(2)}_{14\uparrow}}) \epsilon  - ( j_{12}^2 +  j_{14}^2 ) \epsilon^2  +  \frac{u}{2}(j_{12} \overline{\eta^{(1)}_{12\uparrow}} +  j_{14} \overline{ \eta^{(1)}_{14\uparrow} } ) \epsilon^2 ) \\
    & ~ (  (j_{12}  \overline{\eta^{(2)}_{12\uparrow}} - j_{32}  \overline{\eta^{(2)}_{23\uparrow}}) \epsilon +   ( j_{12}^2 + j_{32}^2 ) \epsilon^2 -   \frac{u}{2} ( j_{12} \overline{\eta^{(1)}_{12\uparrow}} +  j_{32} \overline{\eta^{(1)}_{23\uparrow} }) \epsilon^2  ) \\
    & ~ + O(\epsilon^3) \\
    = & (   j_{12} \overline{\eta^{(2)}_{12 \uparrow}}  - j_{23}  \overline{\eta^{(2)}_{23\uparrow}} ) \epsilon  + ( j_{12}^2 + j_{23}^2 ) \epsilon^2  -  \frac{u}{2} ( j_{12} \overline{\eta^{(1)}_{12\uparrow}} +  j_{32} \overline{\eta^{(1)}_{23\uparrow}} )  \epsilon^2 \\
    & ~ +( - 2 j_{12}^2 \left( \overline{\eta^{(2)}_{12\uparrow}}\right)^2 + 2  j_{12} j_{23} \overline{ \eta^{(2)}_{12\uparrow}} ~ \overline{ \eta^{(2)}_{23\uparrow} }- j_{14}j_{12}\langle \eta^{(2)}_{14\uparrow}\rangle \langle \eta^{(2)}_{12\uparrow} \rangle \\
    & ~ + j_{14}j_{23}\langle \eta^{(2)}_{14\uparrow}\rangle \langle \eta^{(2)}_{23\uparrow} \rangle )\epsilon^2 + O(\epsilon^3) \text{  .}\\
\end{align*}   
\end{scriptsize}
}

Here $\eta_{ij}^{(1)}$ and $\eta_{ij}^{(2)}$ are independent real white noises, so when we take the limit of large number of trajectories $n_{traj}$:

\begin{equation}
  \left\langle \rho_{11\uparrow}+\frac{1}{2} \right\rangle \left\langle  \rho_{22\uparrow}+\frac{1}{2} \right\rangle \xrightarrow[n_{traj} \rightarrow \infty]{}   ( j_{12}^2 + j_{23}^2 ) \epsilon^2  \text{  .}
    \label{eq:g2_den}
\end{equation}
For the nominator of the correlation formula~(\ref{eq:g2_all}), we have from Eqs.~(\ref{eq:ro11u}) and (\ref{eq:ro22u}):
\begin{scriptsize}
\begin{align*} 
   &\langle (\rho_{11\uparrow}+\frac{1}{2})( \rho_{22\uparrow}+\frac{1}{2}) \rangle \\
    =&\langle (1 - ( j_{12} \eta^{(2)}_{12\uparrow} +  j_{14} \eta^{(2)}_{14\uparrow})\epsilon -  (j_{12}^2 + j_{14}^2)\epsilon^2   +  \frac{u}{2}(j_{12} \eta^{(1)}_{12\uparrow} +  j_{14} \eta^{(1)}_{14\uparrow} ) \epsilon^2 )  \\  
    & (( j_{12} \eta^{(2)}_{12\uparrow}  - j_{23}  \eta^{(2)}_{23\uparrow}) \epsilon + (j_{12}^2 + j_{23}^2)\epsilon^2  -  \frac{u}{2} ( j_{12}\eta^{(1)}_{12\uparrow} +  j_{32} \eta^{(1)}_{23\uparrow}) \epsilon^2 ) \rangle \\
    & + O(\epsilon^3)\\
=& \langle  ( j_{12} \eta^{(2)}_{12\uparrow}  - j_{23}  \eta^{(2)}_{23\uparrow}) \epsilon  + ( j_{12}^2 + j_{23}^2 ) \epsilon^2 -   \frac{u}{2} ( j_{12}\eta^{(1)}_{12\uparrow} +  j_{32} \eta^{(1)}_{23\uparrow}) \epsilon^2 \\
&- (  j_{12}^2 (\eta^{(2)}_{12\uparrow})^2 ) \epsilon^2 +  ( j_{12} j_{23} \eta^{(2)}_{12\uparrow} \eta^{(2)}_{23\uparrow} - j_{14}j_{12}\eta^{(2)}_{14}\eta^{(2)}_{12\uparrow} \\
&+ j_{14}j_{23}\eta^{(2)}_{14\uparrow} \eta^{(2)}_{23\uparrow} ) \epsilon^2 \rangle  + O(\epsilon^3) \\
    = & \langle   j_{12} \eta^{(2)}_{12\uparrow}  - j_{23}  \eta^{(2)}_{23\uparrow}\rangle \epsilon  + ( j_{12}^2 + j_{23}^2 ) \epsilon^2  -   \frac{u}{2} ( j_{12}\eta^{(1)}_{12\uparrow} +  j_{32} \eta^{(1)}_{23\uparrow}) \epsilon^2 \\
    & -   j_{12}^2 \langle (\eta^{(2)}_{12\uparrow})^2 \rangle \epsilon^2 +  ( j_{12} j_{23}\langle \eta^{(2)}_{12\uparrow} \eta^{(2)}_{23\uparrow} \rangle - j_{14}j_{12}\langle \eta^{(2)}_{14\uparrow} \eta^{(2)}_{12\uparrow} \rangle \\
    &+ j_{14}j_{23}\langle \eta^{(2)}_{14\uparrow} \eta^{(2)}_{23\uparrow} \rangle ) \epsilon^2  + O(\epsilon^3) \\
=&  ( j_{12}^2 + j_{23}^2 ) \epsilon^2 + \langle   j_{12} \eta^{(2)}_{12\uparrow}  - j_{23}  \eta^{(2)}_{23}\rangle \epsilon - j_{12}^2 \langle (\eta^{(2)}_{12\uparrow})^2 \rangle \epsilon^2 \\
&-   \frac{u}{2} \langle j_{12}\eta^{(1)}_{12\uparrow} +  j_{32} \eta^{(1)}_{23\uparrow} \rangle \epsilon^2 +  ( j_{12} j_{23}\langle \eta^{(2)}_{12\uparrow}\eta^{(2)}_{23\uparrow} \rangle - j_{14}j_{12}\langle \eta^{(2)}_{14\uparrow} \eta^{(2)}_{12\uparrow} \rangle \\
&+ j_{14}j_{23}\langle \eta^{(2)}_{14\uparrow} \eta^{(2)}_{23\uparrow} \rangle ) \epsilon^2 + O(\epsilon^3) \text{  .}\\
\end{align*} 
\end{scriptsize}
Here again, because $\eta_{ij\sigma}^a$ are independent real white noises, we have for the limit of large number of trajectories:
\begin{equation}
\begin{split}
   &\left\langle \left(\rho_{11\uparrow}+\frac{1}{2} \right) \left( \rho_{22\uparrow}+\frac{1}{2} \right) \right\rangle \\
   &\xrightarrow[n_{traj} \rightarrow \infty]{}    ( j_{12}^2 + j_{23}^2 ) \epsilon^2 -  j_{12}^2 \epsilon^2  =  j_{23}^2  \epsilon^2  \text{  .}
\end{split}
    \label{eq:g2_nom}
\end{equation} 
Which, combining (\ref{eq:g2_den}) and (\ref{eq:g2_nom}), gives us:

\begin{equation*}
    g^{(2)}_{12}(\epsilon) \xrightarrow[n_{traj} \rightarrow \infty]{} \frac{ ( j_{23}^2  ) \epsilon ^2}{ (j_{23}^2 + j_{12}^2 ) \epsilon^2} \text{  ,}
\end{equation*}

\begin{equation}
    g^{(2)}_{12}(0) = \frac{ 1 }{1  + \left(\frac{j_{12}}{j_{23}}\right)^2 } \text{  .}
    \label{eq:form_g_st}
\end{equation}

Before we converge the noises $\eta$ to $0$, the leading terms in the fraction~(\ref{eq:g2_all}) are in $O(\epsilon)$. They determine the value of $g^{(2)}(0)$ when the number of trajectories is too small. Hence the difficulties to have an accurate value of $g^{(2)}(0)$, see Fig.~\ref{fig:10p_g2_124}, and the need for more trajectories. 

\subsection{General version of short-time correlation}

In the center of the graphene, the sites have three neighbours. We write a generalisation of Eqs.~(\ref{eq:ro11u}) and (\ref{eq:ro22u}) for an arbitrary number of neighbour sites:

\begin{scriptsize}
\begin{align*}
\rho_{11\uparrow}(\epsilon) &= \frac{1}{2}  - \sum_{i} \left( j_{1i} \eta^{(2)}_{1i\uparrow}  \epsilon  -  j_{1i}^2  \epsilon^2 +  \frac{u}{2} j_{1i} \eta^{(1)}_{1i\uparrow}  \epsilon^2 \right) + O(\epsilon^3) \text{ ,} \\
\rho_{22\uparrow}(\epsilon) &= - \frac{1}{2}  - \sum_{i} \left( j_{2i}  \eta^{(2)}_{2i\uparrow}  \epsilon +    j_{2i}^2  \epsilon^2 -   \frac{u}{2}  j_{2i}\eta^{(1)}_{2i\uparrow} \epsilon^2 \right) + O(\epsilon^3)\text{ .}
\end{align*}
\end{scriptsize}
Which leads to a generalisation of Eq.~(\ref{eq:form_g_st}) 


\begin{equation}
    g^{(2)}_{12}(0) = \frac{ (\sum_{i}{j^2_{2i}}) - j^2_{12} }{\sum_{i}{j^2_{2i}} } \text{  .}
\label{eq:form_g_st_2}
\end{equation}
If the site $2$ has three neighbours (or in the case of Fig.~\ref{fig:198p_g2_pos} the site 36), and all hopping interactions $j$ are equals, we find $g^{(2)}_{12}(0) = 2/3$, see Fig.~\ref{fig:198p_g2_neib}.

\subsection{Large-time correlation functions}
\label{sec:longt_corr}

The correlation between site $i$ and site $j$, $g^{(2)}_{ij}$, can be understood as the effect of the knowledge of the presence of a particle on site $i$ on the probability to find a particle on site $j$:

\begin{equation}
    g^{(2)}_{i,j} = \frac{P(n_i = 1 | n_j = 1)}{P(n_i = 1)}\text{  .}
    \label{eq:g2_prob}
\end{equation}
For $p$ particles in a Fermi-Hubbard system of $n$ sites, when the probability density is totally spread on all sites, we express the probability with the binomial coefficients $C^p_n = \frac{n !}{p! (n-p)!}$:
\begin{equation}
    P(n_j = 1) = \frac{C^{p-1}_{n-1}}{C^{p}_{n}}\text{  ,}~~ P(n_j = 1|n_i = 1) = \frac{C^{p-2}_{n-2}}{C^{p-1}_{n-1}}\text{  .}
    \label{eq:form_g_lt_1}
\end{equation}

For a half-filled system, $p = n/2$,  $P(n_j = 1) = 1/2$, such that Eq.~(\ref{eq:g2_prob}) becomes,

\begin{equation}
    g^{(2)}_{i,j} = 2 \frac{C^{n/2-2}_{n-2}}{C^{n/2-1}_{n-1}} \text{  .}
    \label{eq:form_g_lt_2}
\end{equation}
We give in table~\ref{tab:g2_res} some examples of results from~(\ref{eq:form_g_lt_2}) for systems of different sizes, and compare it with numerical results in Fig.~\ref{fig:np_g2}.

\vspace{1cm}
\begin{table}[h!]
\centering
\begin{tabular}{ c | c c c c |}
 $n$ & 4 & 6 & 10 & 198 \\ 
 \hline
 $g^{(2)}_{i,j}$ & $\frac{2}{3}$ & $0.8$ & $\frac{8}{9}$ & 0.995 
\end{tabular}
 \caption{Limit ($t\rightarrow \infty$) values of correlation function for system of different sizes $n$.}
  \label{tab:g2_res}
\end{table}

\vspace{4cm}

\end{document}